\documentclass[aps,prl,article,twocolumn,citeautoscript,footinbib,superscriptaddress]{revtex4-2} \synctex=1 
\usepackage{amsmath,amssymb,bm,bbm} 
\usepackage{amsthm}
\usepackage{physics}
\usepackage{graphicx}  

\usepackage{amsmath,amsfonts,tikz}
\usepackage{graphicx,xcolor,multirow}
\usepackage[colorlinks=true]{hyperref}  
\usepackage[section]{placeins}
\usepackage[mathscr]{euscript}
\usepackage[toc,page]{appendix}
\hypersetup{
    unicode=false,          
    pdftoolbar=true,        
    pdfmenubar=true,        
    pdffitwindow=false,     
    pdfstartview={FitH},    
    pdftitle={},    
    pdfauthor={},     
    pdfsubject={},   
    pdfcreator={},   
    pdfproducer={}, 
    pdfkeywords={} {} {}, 
    pdfnewwindow=true,      
    colorlinks=true,       
    linkcolor=magenta, 
    citecolor=blue,        
    filecolor=magenta,      
    urlcolor=blue           
} 

\usepackage{pdfpages}
\usepackage{pgffor}
\makeatletter
\AtBeginDocument{\let\LS@rot\@undefined}
\makeatother

\usepackage{pifont}

\newcommand{\La}{\line (1,0  ){12}}
\newcommand{\Lb}{\line (3,5 ){6}}

\newcommand{\Ld}{\line (-1,0){12}}
\newcommand{\Le}{\line (-3,-5){6}}

\newcommand{\C} {\circle*{4}}
\newcommand{\Ll}{\line (-1,0  ){1}}
\newcommand{\LlT}{\rule[-1pt]{0.5cm}{0.2em}}  

\newcommand{\LaT}{\rule[-1pt]{0.4cm}{0.2em}}  
\newcommand{\LdT}{\rule[-1pt]{0.4cm}{0.2em}}  
\newcommand{\LbT}{\rotatebox{60}{\rule[-1pt]{0.4cm}{0.2em}}}  
\newcommand{\LeT}{\rotatebox{60}{\rule[-1pt]{0.4cm}{0.2em}}}  

\newcommand{\pA}{\put(-6,-10)}
\newcommand{\pB}{\put(6,-10)}
\newcommand{\pC}{\put(12,0)}

\newcommand{\pZ}{\put(0,0)}

\newcommand{\pM}{\put(-5,-3)}

\newcommand{\pAT}{\put(-6.8,-10)} 
\newcommand{\pBT}{\put(5.2,-10)}  

\newcommand{\rhomb}{
  \pA{\C}\pB{\C}\pZ{\C}\pC{\C}
 }

\newcommand{\rhombH}{
  \begin{picture}(22,10)(-8,-6)
    \pA{\LaT}\pB{\Lb}\pZ{\Le}\pZ{\LdT}
    \rhomb
  \end{picture}
}
\newcommand{\rhombV}{
  \begin{picture}(22,10)(-8,-6)
   \pA{\La}\pBT{\LbT}\pAT{\LeT}\pC{\Ld}
    \rhomb
  \end{picture}
}
\newcommand{\onedimer}{
  \begin{picture}(22,10)(-8,-6)
\pM{\C}
\pM{\LlT\C}
  \end{picture}
}

\newcommand{\onelink}{
  \begin{picture}(22,10)(-8,-6)
        \pM{\C \Ll \La \C}
  \end{picture}
}

\usepackage{amsfonts}
\usepackage{upgreek}
\usepackage{slashed}
\usepackage{latexsym}
\usepackage{mathrsfs}
\usepackage{tikz-feynman}
\usepackage[usestackEOL]{stackengine}
\usepackage{natbib} 
\begin{document}
\title{Triangular lattice quantum dimer model with variable dimer density}

\author{Zheng Yan}
\affiliation{Department of Physics and HKU-UCAS Joint Institute of Theoretical and Computational Physics, The University of Hong Kong, Pokfulam Road, Hong Kong SAR, China}

\author{Rhine Samajdar}
\affiliation{Department of Physics, Harvard University, Cambridge, MA 02138, USA}

\author{Yan-Cheng Wang}
\affiliation{Beihang Hangzhou Innovation Institute Yuhang, Hangzhou 310023, China}

\author{Subir Sachdev}
\email{sachdev@g.harvard.edu}
\affiliation{Department of Physics, Harvard University, Cambridge, MA 02138, USA}
\affiliation{School of Natural Sciences, Institute for Advanced Study, Princeton, NJ 08540, USA}

\author{Zi Yang Meng}
\email{zymeng@hku.hk}
\affiliation{Department of Physics and HKU-UCAS Joint Institute of Theoretical and Computational Physics, The University of Hong Kong, Pokfulam Road, Hong Kong SAR, China}

\date{\today}

\begin{abstract}
\noindent{\bf Abstract} Quantum dimer models are known to host topological quantum spin liquid phases, and it has recently become possible to simulate such models with Rydberg atoms trapped in arrays of optical tweezers. Here, we present large-scale quantum Monte Carlo simulation results on an extension of the triangular lattice quantum dimer model with terms in the Hamiltonian annihilating and creating single dimers. We find distinct odd and even $\mathbb{Z}_2$ spin liquids, along with several phases with no topological order: a staggered crystal, a nematic phase, and a trivial symmetric phase with no obvious broken symmetry. We also present dynamic spectra of the phases, and note implications for experiments on Rydberg atoms.
\end{abstract}

\maketitle

\vspace{\baselineskip}
\noindent{\bf Introduction}

Recent quantum simulation advances have provided remarkable microscopic access to the quantum correlations of a $\mathbb{Z}_2$ quantum spin liquid (QSL) \cite{Roushan21,Semeghini21}. 
The $\mathbb{Z}_2$ QSL \cite{RS91,wen1991} is the simplest quantum state in two spatial dimensions with fractionalized excitations and time-reversal symmetry, and has the same anyon content as the toric code \cite{Kitaev1997}. Once we include considerations of lattice and other symmetries,  $\mathbb{Z}_2$ QSLs come in different varieties; the distinctions between them are important in understanding the phase diagrams of possible experimental realizations. The coarsest classification subdivides $\mathbb{Z}_2$ QSLs into `odd' and `even' classes, depending upon whether elementary translations anti-commute or commute when acting on excitations carrying $\mathbb{Z}_2$ magnetic flux \cite{RJSS91,MVSS99,TSMPAF99,Moessner01}, and results in different translational symmetry fractionalization patterns and spectral signatures in the dynamic response \cite{Essin2014,JWMei2015,GYSun2018,YCWang2021NC,YCWang2017QSL,YCWang2018}. More refined classifications have been obtained since \cite{Essin:2013rca,Zaletel:2014epa,Cheng:2015kce,QiMeng16,Bulmash:2020flp}.

Quantum dimer models (QDMs) \cite{RK88,moessner2011quantum} on nonbipartite lattices have long been known to host $\mathbb{Z}_2$ QSLs. 
In this work, we investigate an important---but hitherto unexplored---extension of the quantum dimer model  on the triangular lattice \cite{RMSLS01,roychowdhury2015z,Plat2015z2}. Unlike the more conventionally studied QDMs, here, the density of dimers is allowed to vary by terms in the Hamiltonian which can annihilate and create single dimers on each link of the triangular lattice. Such a dimer-nonconserving term is motivated by connections to models of ultracold atoms trapped in optical tweezers \cite{fendley2004competing,Bernien17}, in which each dimer is identified with an atom excited to a Rydberg state by laser pumping \cite{Samajdar:2020hsw,Verresen:2020dmk, Z2GT}. 
The observations of \citet{Semeghini21} are for the case where the atoms are positioned on the {\it links\/} of the kagome lattice; this connects to the quantum dimer model on the \textit{kagome} lattice \cite{Verresen:2020dmk}. Our study pertains to the \textit{triangular} lattice dimer model, which connects to the case where the atoms are placed on the {\it sites\/} of the kagome lattice \cite{roychowdhury2015z,Plat2015z2,Samajdar:2020hsw}; such a configuration can be readily realized in the experiments, and initial explorations of quantum phases in such a lattice have already been carried out by the team of Ref.\cite{Semeghini21}.

With a dimer-nonconserving term present, here we show, the triangular lattice quantum dimer model displays novel features relevant to the Rydberg-atom experiments. When the nonconserving terms are large, we can obtain a `trivial' phase with neither topological order nor broken lattice symmetry. More interestingly, the phase diagram of this extended QDM also harbors both odd and even $\mathbb{Z}_2$ liquids. Note that in early discussions of such QSLs in dimer models, the distinction between the liquids was tied to whether the number of dimers on each site was constrained to be odd or even~\cite{roychowdhury2015z,Plat2015z2}. In the present model, the number of dimers on each site fluctuates between odd and even values, namely $1$ and $2$; nevertheless, the distinction between even and odd QSLs still survives based on the symmetry transformation properties of excitations with magnetic $\mathbb{Z}_2$ flux (`visons'). In the case with a dimer number constraint on each site, there is an anomaly relation requiring that odd (even) dimers produce vison translations which anticommute (commute)~\cite{Cheng:2015kce,QiMeng16}. However, in the case without a dimer number constraint (or a soft constraint), of interest to us here, microscopic details will determine whether vison translations anticommute or commute, and we will investigate this fate numerically with quantum Monte Carlo simulations.

Finally, our study also obtains several phases which break lattice symmetries, but are topologically trivial. This includes two `staggered' phases \cite{RMSLS01}, a `columnar' phase~\cite{Ralko2005} and a `nematic' phase \cite{fradkin14,roychowdhury2015z,Plat2015z2}, and we also discuss their density-wave-ordered counterparts in the context of experiments on Rydberg quantum simulators.
\begin{figure*}[tp!]
\centering
\includegraphics[width=2\columnwidth]{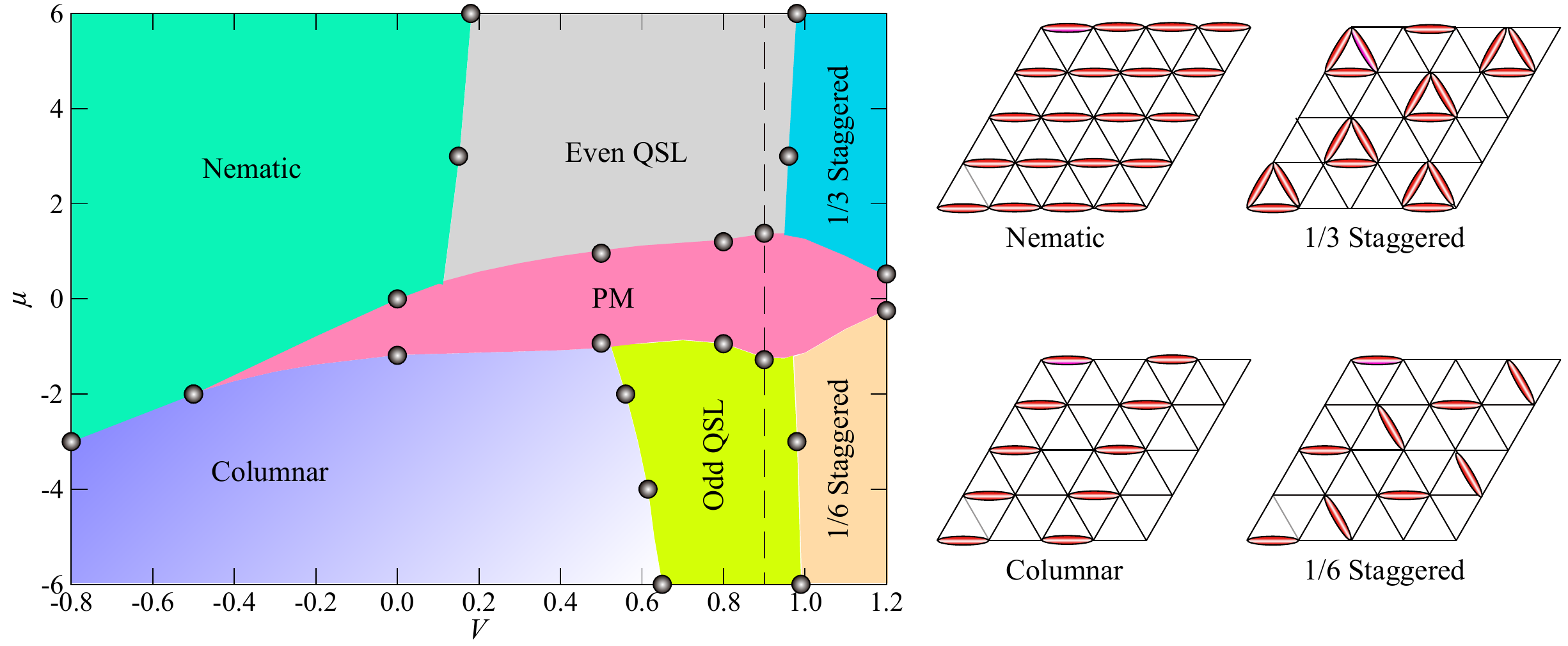}
\caption{\textbf{Phases of the variable-density triangular lattice QDM.} \textsl{Left panel:} The full phase diagram, spanned by the $V$ and $\mu$ axes, is obtained from QMC simulations at $h=0.4$. The phase boundaries between the paramagnetic (PM) phase and the two QSLs along the dashed line are studied in Fig.~\ref{Fig3}; the phase transitions are first-order. The phase boundaries between the QSLs and the nematic, columnar, and staggered phases are shown in Supplementary Note~3 of the Supplementary Information (SI). The associated transitions are either continuous (such as the QSL--nematic and QSL--columnar) or first-order (such as the QSL--staggered). 
\textsl{Right panel:} Schematic pictures of the four crystalline phases (nematic, columnar, $1/3$ staggered, and 1/6 staggered). In the limit of exactly one dimer per site, a $\sqrt{12}\times\sqrt{12}$ valence bond solid (VBS) phase is known to exist between the odd QSL and the columnar phase. However, it is nearly degenerate with the columnar phase over a large region in our simulations, and we depict this schematically by using a lighter shading for the columnar phase near the odd QSL.}
    \label{Fig1}
\end{figure*}

\vspace{\baselineskip}
\noindent{\bf Results}

\noindent{\bf The model.} We investigate the following general dimer Hamiltonian, with one or two dimer(s) per site, on the triangular lattice,
\begin{eqnarray}
  H=&-t&\sum_r \left(
  \left|\rhombV\right>\left<\rhombH\right| + \mbox{h.c.}
  \right) \nonumber \\
  &+V&\sum_r\left(
  \left|\rhombV\right>\left<\rhombV\right|+\left|\rhombH\right>\left<\rhombH\right|
  \right)\nonumber \\
  &-h& \sum_l \left(\left|\onedimer\right>\left<\onelink\right|+\mbox{h.c.}\right)\nonumber \\
  &-\mu& \sum_l \left(\left|\onedimer\right>\left<\onedimer\right|\right),
\label{eq:eq1}
\end{eqnarray}
where the sum on $r$ runs over all plaquettes (rhombi), including the three possible
orientations, and $l$ runs over all links. The different terms in this Hamiltonian are as follows. The kinetic term (controlled by $t$) flips the
two dimers on every flippable plaquette, {\it i.e.\/}, on each plaquette with two
parallel dimers, while the potential term (controlled
by the interaction $V$) describes a repulsion ($V>0$) or an attraction ($V<0$)
between nearest-neighbor dimers. The transverse-field term of strength $h$ creates/annihilates a dimer at link $l$ (similar terms also appear in the quantum realization of the classical models of Ref.~\onlinecite{fradkin14}), in contrast to the $t$ and $V$ terms, neither of which change the dimer number. Lastly, $\mu$ sets the chemical potential for the occupation of a link by a dimer. We further impose a soft constraint requiring that there must be one or two dimer(s) per site. 
Thus, when $\mu \to \pm\infty$, the model reverts to the conventional hard-constrained quantum dimer model with exactly two or one dimer(s) per site---the phase diagrams of both these QDMs have been extensively studied in the literature~\cite{MoessnerSondhi2001a,MoessnerSondhi2001b,Ralko2005,Ralko2006,Ralko2008,roychowdhury2015z,Plat2015z2,ZY2020}. Hereafter, we set $t=1$ as the unit of energy for the rest of this paper.

\begin{figure*}[htp]
\centering
\includegraphics[width=2\columnwidth]{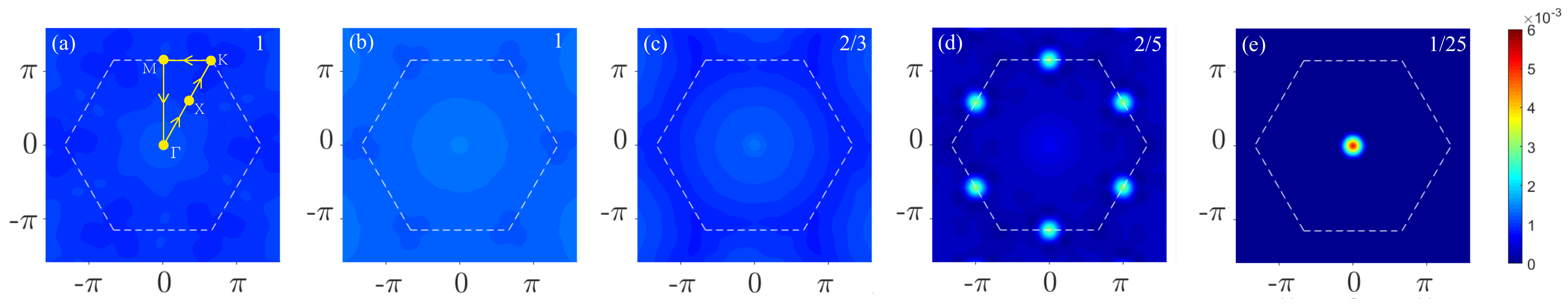}
\caption{\textbf{Equal-time dimer structure factors.} Here, we present $D(\mathbf{k},\tau=0)$ in the Brillouin zone for the (a) odd $\mathbb{Z}_2$ QSL ($\mu=-3$, $V=0.9$), (b) PM phase ($\mu=0$, $V=0.9$), (c) even $\mathbb{Z}_2$ QSL ($\mu=3$, $V=0.9$), (d) columnar phase ($\mu=-3$, $V=-0.5$), and (e) nematic phase ($\mu=3$, $V=-0.5$) in the phase diagram of Fig.~\ref{Fig1}. All the data are simulated using $\beta=L=12$. The upper-right labels in each panel represent the scaling factor for the intensities such that the five panels can be scaled onto the same colorbar. In addition, the high-symmetry path for the spectra in Fig.\ref{Fig4} are also drawn in (a).}
\label{Fig2}
\end{figure*}

To solve the model in Eq.~\eqref{eq:eq1} in an unbiased manner, we employ the recently developed sweeping cluster quantum Monte Carlo algorithm, which can perform efficient sampling in constrained quantum many-body systems~\cite{ZY2019sweeping,ZY2020improved,ZY2020,ZY2021mixed}. By monitoring the behavior of various physical observables such as dimer correlation functions and structure factors, we map out the detailed phase diagrams, such as, for instance, in Fig.~\ref{Fig1}. Moreover, in addition to static observables, we also compute the dynamic dimer correlation functions in imaginary time and employ the stochastic analytic continuation method~\cite{Sandvik1998a,Beach2004,Shao2017nearly,GYSun2018,ZY2020,Zhou2021amplitude,YCWang2021vestigial,YCWang2021NC,shao2022progress} to obtain the dynamic dimer spectral functions in real frequencies. Our simulations are performed on the triangular lattice with periodic boundary conditions and system sizes $N=3L^2$ for linear dimensions $L=8,12,16,18,24$, while setting the inverse temperature $\beta=L$ ($\beta=200$) for equal-time (dynamical) simulations.

\noindent{\bf The phase diagram.}
Although the phase diagrams in the two limits with exactly $1/3$ and $1/6$ dimer fillings are well understood, the manner in which they connect to each other in the presence of a nonzero transverse field $h$ and chemical potential $\mu$ is an interesting open question. In particular, one may ask what happens between the two kinds of $\mathbb{Z}_2$ QSLs, {\it i.e.\/}, whether they are separated by a direct phase transition or an intermediate phase. An important reason this question has remained unaddressed so far is the lack of a suitable algorithm to deal with the soft constraint. As discussed in detail in the section of method, here, we adapt the sweeping cluster Monte Carlo algorithm used for hard-constrained QDMs~\cite{ZY2019sweeping,ZY2020improved} to soft ones and use it to map out the phase diagram of the Hamiltonian in Eq.~\eqref{eq:eq1}. Figure~\ref{Fig1} shows the full phase diagram obtained at $h=0.4$, which we focus on in the main text, leaving the discussion of similar phase diagrams with different $h$ to Supplementary Notes~2, 3 of the Supplementary Information (SI).

The phase diagram exhibits four different symmetry-breaking phases, including the nematic, the columnar, and two staggered phases; the schematic plots of these crystalline phases are shown in the right panels of Fig.~\ref{Fig1}. Furthermore, we observe two distinct $\mathbb{Z}_2$ QSL phases, which are denoted as `Even QSL' and `Odd QSL' in the figure. Additionally, a trivial disordered---or paramagnetic (PM)---phase exists in the central region in between the two QSLs; note that such a PM phase does not arise in the more familiar QDMs where the dimer number per site is exactly constrained. The phase boundaries between these phases are determined by examining various parameter points and paths scanning through the phase diagram, such as the dashed line in Fig.~\ref{Fig1}.

To characterize this rich variety of phases, we compute the equal-time $(\tau$\,$=$\,$0)$ dimer structure factor (see Fig.~\ref{Fig2}) as
\begin{equation}
D(\mathbf{k},\tau)=\frac{1}{N}
\hspace*{-0.15cm}
\sum_{\substack{i,j \\ \alpha=1,2,3}}^{L^3}
\hspace*{-0.15cm}
e^{i\mathbf{k}\cdot \mathbf{r}_{ij}} \left(\langle n^{}_{i,\alpha}(\tau)n^{}_{j,\alpha}(0)\rangle - \langle n^{}_{i,\alpha}\rangle \langle n^{}_{j,\alpha}\rangle \right),
\end{equation}
where $n_i$ is the dimer number operator on bond $i$ and $\alpha$ stands for the three bond orientations, at five representative parameter points corresponding to the five different phases in the phase diagram. Figures~\ref{Fig2}(a), (b), and (c) show $D(\mathbf{k},0)$ inside the odd QSL, PM, and even QSL phases, respectively. In the hexagonal Brillouin zone, we observe that there are no peaks associated with long-range order but only broad profiles signifying different short-range dimer correlation patterns in real space. In contrast, Figs.~\ref{Fig2}(d) and (e) present the dimer structure factors inside the columnar and nematic phases, respectively. One now clearly sees the Bragg peaks at the $M$ points for the columnar phase (there can be three different orientations of the columnar dimers, corresponding to all the 3 pairs of $M$ points), and at the $\Gamma$ point in the nematic phase.

\begin{figure}[htp!]
\centering
\includegraphics[width=\columnwidth]{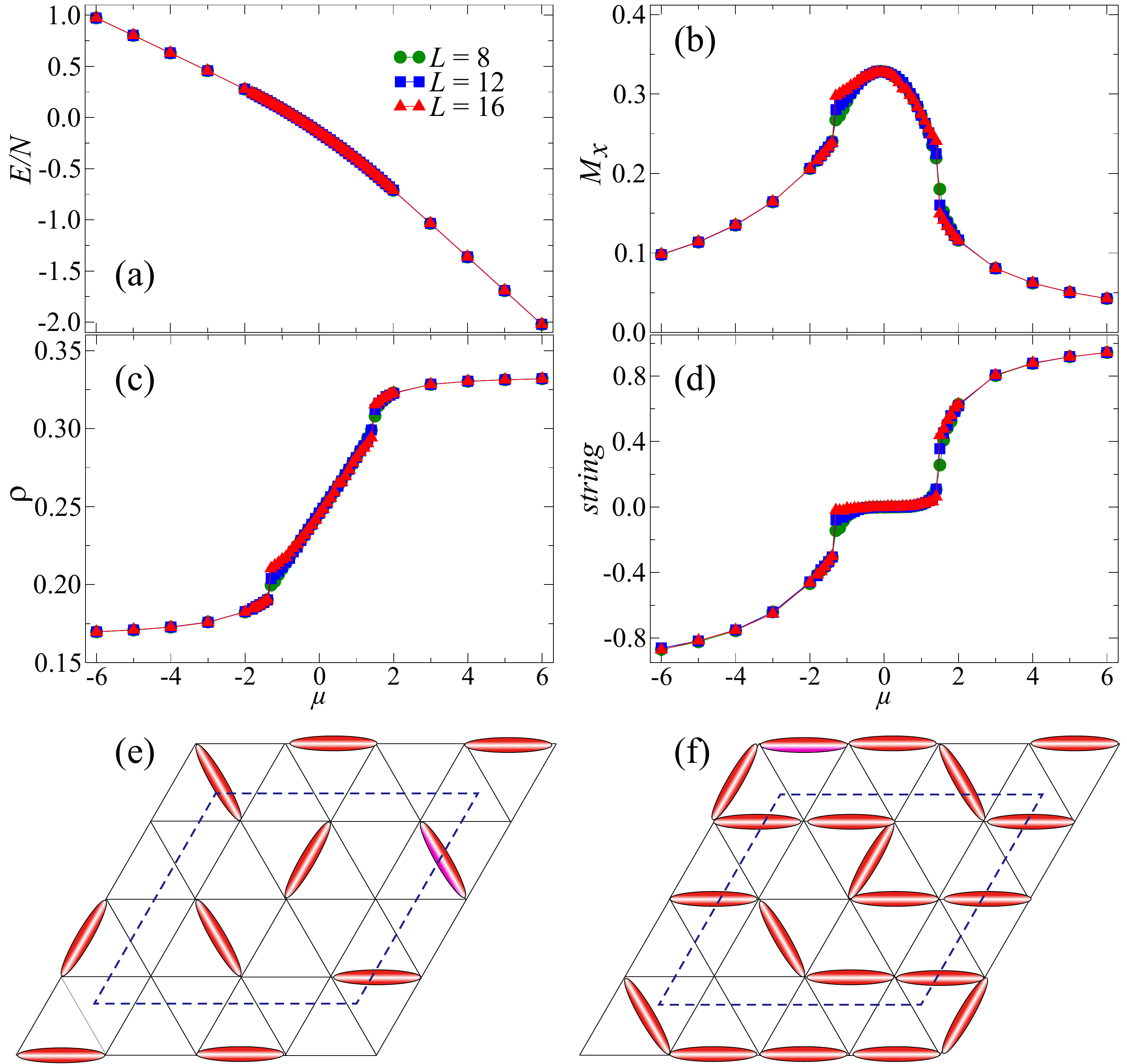}
\caption{\textbf{Phase transitions between QSLs and the PM phase.} Data along the QSL--PM--QSL path, indicated by the dashed line at $V=0.9$ in Fig.~\ref{Fig1}. (a) The energy density is smooth with increasing $\mu$. (b) The polarization $M_x$ reveals the first-order phase transition between the PM phase and the two $\mathbb{Z}_2$ QSLs. (c) The dimer filling remains at approximately $\rho \sim 1/3$ in the even QSL and $\rho \sim 1/6$ in the odd QSL. It changes continuously in the PM phase, and the filling also exhibits a first-order phase transition between the PM phase and QSLs. (d) The string operator is zero in the trivial PM phase but  positive (negative) in the even (odd) $\mathbb{Z}_2$ QSL. All the data are calculated for $V=0.9$, $\beta=L$, $h=0.4$. (e) In a pure odd $\mathbb{Z}_2$ QSL with dimer filling $\rho=1/6$, a string operator defined on a rhomboid with odd linear size (3 in this case) should attain the value $-1$. (f) In a pure even $\mathbb{Z}_2$ QSL with dimer filling $\rho=1/3$, the string operator should always yield $1$. The string operators presented in (d) are measured for a $3\times 3$ rhombus averaged over the entire lattice for different $L$.}
    \label{Fig3}
\end{figure}

\begin{figure*}[htp!]
\centering
\includegraphics[width=2\columnwidth]{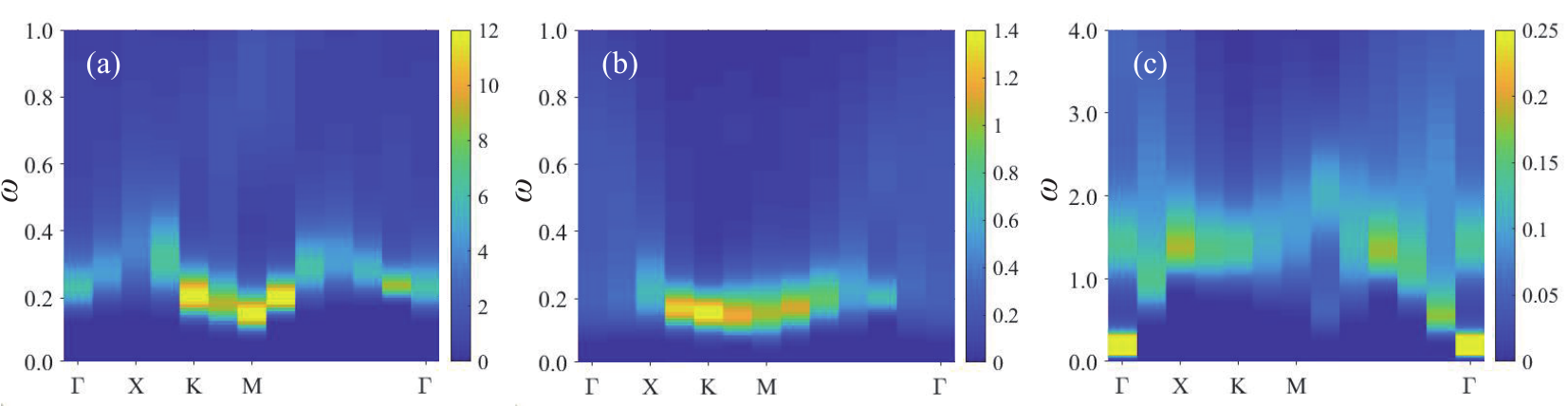}
\caption{\textbf{Dynamical dimer spectra.} The dimer spectra in the (a) odd $\mathbb{Z}_2$ QSL in the limit of one dimer per site, corresponding to $\mu\to -\infty$ and $V=1$ in Fig.~\ref{Fig1}, (b) PM phase with  $\mu=0, V=0.9$ and $h=0.4$, and (c) even $\mathbb{Z}_2$ QSL in the limit of two dimers per site, corresponding to $\mu \to \infty$ and $V=0.5$ in Fig.~\ref{Fig1}. The dimer spectra exhibit continua in both (a) and (c), conveying the fractionalization of spins into visons. However, the dispersion minima in the two cases differ, being located at both $M$ and $\Gamma$ for (a) and only at $\Gamma$ for (c), representing the translational symmetry fractionalization in the former and the lack thereof in the latter. In (b), however, the dimer spectrum is flat and displays less of a continuum in the frequency domain, consistent with a polarized PM phase. All the data are simulated at $\beta=200$ on a $L=12$ lattice, with the low temperature $T=1/200$ being necessary to overcome the small vison gap and the transverse field $h$.}
\label{Fig4}
\end{figure*}

\noindent{\bf The two $\mathbb{Z}_2$ QSLs.}
Having established the lack of long-range dimer-dimer correlations in the odd/even $\mathbb{Z}_2$ QSLs and the PM phase, next, we move on to the phase transitions between them. Since all three of these phases are disordered, care needs to be taken in determining their phase boundaries. Our results in this regard are summarized in Fig.~\ref{Fig3}, which shows the data along a path with a fixed $V=0.9$ and varying $\mu$ in the phase diagram (dashed line in Fig.~\ref{Fig1}). 

First, in Fig.~\ref{Fig3}(a), we illustrate the energy density curves, which appear to be smooth without any obvious turning points along the path as $\mu$ is scanned. However, when the transverse field becomes large, we expect that all the links should be polarized along the $x$ axis (if there were no constraints). Since the model in Eq.~\eqref{eq:eq1} can be regarded as a spin model with spins on links (occupied/empty links being equivalent to spin up/down), the polarization
\begin{equation*}
M_x=\frac{1}{N}\sum_l \left(\left|\onedimer\right>\left<\onelink\right|+\mbox{h.c.}\right) \sim \frac{1}{N}\sum_l S^x_l    
\end{equation*}
can be used to describe the level of polarized links (spins), and thus, to probe the PM phase. Indeed, as seen in Fig.~\ref{Fig3}(b), $M_x$ helps us to identify a first-order phase transition between the PM phase and the two $\mathbb{Z}_2$ QSLs. In the PM phase, $M_x$ becomes large but is still far from the classical saturation value of $1$; this is because the soft constraint forbids all links from being fully polarized simultaneously.
We can also discover similar first-order phase transitions, at the same parameter points, independently from the dimer filling $\rho$ shown in Fig.~\ref{Fig3}(c). In the even (odd) $\mathbb{Z}_2$ QSL phase, the filling is nearly $1/3$ ($1/6$) while the filling changes continuously in the PM phase. 

Additionally, a closed string operator \cite{Semeghini21}, schematically defined as in Figs.~\ref{Fig3}(e) and (f) as $\langle string \rangle= \langle (-1)^{\#~\mathrm{cut~dimers}} \rangle$ on a rhomboid with odd linear size, can be used to distinguish the two QSLs and the PM phase. As shown in Figs.~\ref{Fig3}(e) and (f), $\langle string \rangle$ should be $\pm 1$ in a \textit{pure} even/odd $\mathbb{Z}_2$ QSL without spinons and $0$ in a PM phase. We measure all the $3\times 3$ rhomboids in the lattice to obtain the expectation value $\langle string \rangle$ along the path scanning $\mu$ at $V=0.9$. The resultant data in Fig.~\ref{Fig3}(d) indeed reveal that inside the odd (even) $\mathbb{Z}_2$ QSL phase, $\langle string \rangle$\,$\sim$\,$-1$ ($\langle string \rangle$\,$\sim$\,$1$), while inside the PM phase, $\langle string \rangle$\,$\approx$\,$0$; the transitions are also seen to be first-order, in consistency with Figs.~\ref{Fig3}(b) and (c).

\noindent{\bf The dynamical dimer spectra.}
One of the hallmarks of a QSL is its ability to support fractionalized excitations that cannot be created individually by any local operator. In this section, we focus on one class of such fractional excitations with magnetic $\mathbb{Z}_2$ flux, {\it i.e.\/}, the visons. Naturally, vison configurations with different fluxes will result in different dimer spectral signatures, thus realizing, in particular, the interesting phenomenon of translational symmetry fractionalization~\cite{Essin2014,JWMei2015,GYSun2018,YCWang2021NC,ZY2020}, which can be further used to distinguish the PM and the even/odd $\mathbb{Z}_2$ QSLs and make possible connection to experiments. To this end, we compute the dimer spectra, obtained from stochastic analytic continuation of the Monte-Carlo-averaged dynamic dimer correlation function $D(\mathbf{k},\tau)$ with $\tau \in [0,\beta]$ (which can be viewed as the dynamical vison-pair correlation functions deep inside the $\mathbb{Z}_2$ QSLs~\cite{ZY2020}; more details can be found in the Supplementary Note 1). Figure~\ref{Fig4}(a) shows that in the odd $\mathbb{Z}_2$ QSL phase, the gapped dimer (vison-pair) spectrum forms a continuum, and the dispersion minima are located at both the $M$ and $\Gamma$ points~\cite{Ralko2006,ZY2020}. On the other hand, Fig.~\ref{Fig4}(c) illustrates that the dimer (vison-pair) spectrum deep inside the even $\mathbb{Z}_2$ QSL is also a continuum but with minima only at $\Gamma$. These features are consistent with the expectation that the visons of the odd $\mathbb{Z}_2$ QSL carry a fractional crystal momentum whereas visons of the even QSL do not~\cite{GYSun2018,ZY2020}. Note that for the single vison dispersion of an odd QSL, the locations of the minima are dependent on the chosen gauge~\cite{Ivanov2004PRB,Ralko2007PRB, Z2GT} whereas the vison-pair spectrum is a gauge-invariant observable. For the even QSL, Refs.~\onlinecite{roychowdhury2015z,Plat2015z2} found that the minima of the mean-field vison dispersion occur at the three inequivalent $M$ points in the Brillouin zone. Accordingly, one would then expect the vison-\textit{pair} spectrum to exhibit a minimum at $\Gamma$ (which is equivalent to $2 M$ modulo a reciprocal lattice vector), in agreement with our numerical results. The arguments above apply generally to the dynamics of an odd/even QSL and should hold even at finite $\mu$; similar behaviors have also been observed for the odd/even QSLs of the Balents-Fisher-Girvin (BFG) model~\cite{GYSun2018}. In comparison, Fig.~\ref{Fig4}(b) presents the dimer spectrum inside the PM phase; here, there exists no clear continuum in the frequency domain, indicating the lack of fractionalization of dimers into pairs of visons. Moreover, the overall dispersion is flat, which is consistent with the dispersionless $S^{z}$ spectrum in an $S^{x}$-polarized state, such as in the transverse-field Ising model.

\vspace{\baselineskip}
\noindent{\bf Discussion}

In this work, we investigate a QDM with variable dimer density on the triangular lattice and uncover a plethora of interesting phases, including crystalline solids and two distinct classes of highly entangled QSL states hosting fractionalized excitations. Through detailed quantum Monte Carlo analyses, we explore the subtle interplay between these different phases and find the unique properties of their static and dynamic fingerprints. With the remarkable advances in quantum simulation, experimental realization of the dimer model in Eq.~\eqref{eq:eq1} should provide new probes of novel QSL phases and their phase transitions.

In particular, our results could find application to recent experiments with programmable quantum simulators based on highly tunable Rydberg atom arrays, which have emerged as powerful platforms to study strongly correlated phases of matter and their dynamics. 
While our extended QDM differs from models of Rydberg atoms on the sites of the kagome lattice \cite{Samajdar:2020hsw} in the precise form of the $V$ interactions, the two systems bear resemblance in some of their phases. 
Specifically, the Rydberg model also displays the 1/6 staggered and nematic phases of Fig.~\ref{Fig1}, separated by a `liquid' regime with no broken symmetry. These ordered phases can be mapped to the solid phases of a triangular-lattice QDM with either one or two dimers per site, which precisely constitutes our soft constraint.  
Appealing to the universality of phase transitions \cite{Samajdar:2020hsw}, possible fates of the liquid state in the Rydberg model are then one or more of the phases obtained by interpolating between the $1/6$ staggered and nematic phases in Fig.~\ref{Fig1} for the present quantum dimer model: namely, the odd QSL, the PM, and the even QSL. These considerations highlight the potential utility of variable-density dimer models in the experimental realm and provide a pathway to studying their rich physics.

\vspace{\baselineskip}
\noindent{\bf Methods}\\
\noindent{\bf Sweeping cluster algorithm.} 
This is a quantum Monte Carlo method developed by the authors to solve the path integral of constrained quantum many-body models~\cite{ZY2019sweeping,ZY2020improved,ZY2020,ZY2021mixed,ZY2022hQDM}. The key idea of the sweeping cluster algorithm is to sweep and update layer by layer along the imaginary-time direction, so that the local constraints (gauge fields) are recorded by update lines. In this way, all the samplings are performed in the restricted Hilbert space, \textit{i.e.}, the low-energy space. 
The original sweeping cluster QMC method~\cite{ZY2019sweeping,ZY2020improved} is designed for hard-constraint models, \textit{i.e.}, models in which the number of dimer(s) per site is fixed~\cite{ZY2020,ZY2022loop}. To solve our models in this work, we further improve upon the prior methods to be able to simulate a soft-constrained dimer model.

The Hamiltonian that we consider is given by Eq.~\eqref{eq:eq1}
supplemented with the `soft' constraint that there can only be either one or two dimer(s) per site. The definition of winding numbers~\cite{ZY2020improved,zhou2020quantumstring,zhou2020quantum,yan2022targeting,zhou2021emergent} for these two cases are explained in the Supplementary Note 3.

\begin{figure*}[htp]
\centering
    \includegraphics[width=2\columnwidth]{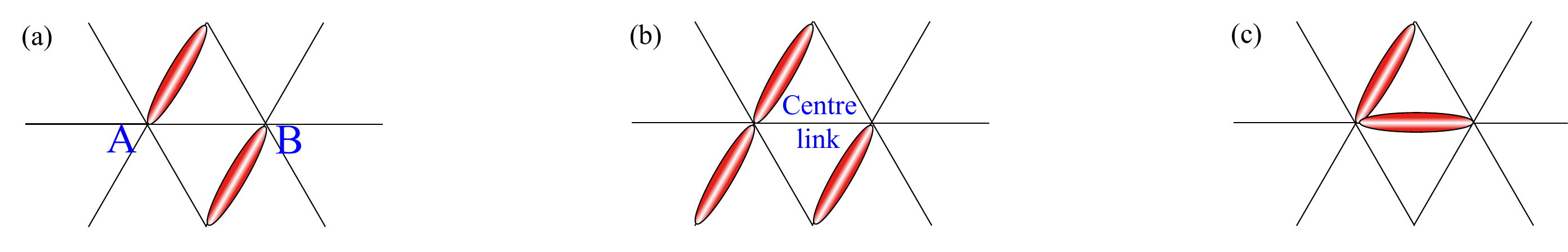}
    \caption{\textbf{Update scheme of soft constraint.} For the soft constraint of $1$ or $2$ dimer(s) per site, we have to consider all the neighbors when creating/annihilating a dimer on the central link. (a) When both the A and B sites have one dimer, one is allowed to create/annihilate a dimer on the centre link. (b) It is forbidden to create a dimer on the centre link when either the A or the B site already has two dimers. (c) It is forbidden to annihilate a dimer on the centre link when either the A or the B site has only one dimer.}
    \label{SF1}
\end{figure*}

Similar to the practice in Stochastic Series Expansion types of quantum Monte Carlo methods~\cite{Sandvik2010b}, we separate the Hamiltonian into diagonal and off-diagonal parts. It is obvious that the $t$ and $V$ terms will not change the number of dimer(s) per site, but both the chemical potential $\mu$ and the transverse field term $h$ would. Therefore, the Monte Carlo update will need to obey the soft constraint when we deal with the $\mu$ and $h$ terms. We write the $h$ off-diagonal term and the $\mu$ diagonal term as,
\begin{eqnarray}
  H^{}_{d,l}&=&\mu\left(\left|\onedimer\right>\left<\onedimer\right|\right)+C, \\ 
  H^{}_{o,l}&=& h \left(\left|\onedimer\right>\left<\onelink\right|+\mbox{h.c.}\right),  
\label{eq:eq2}
\end{eqnarray}
where $C$ is a constant to ensure that the corresponding matrix elements are positive. The label $``d/o"$ indicates whether the operator is diagonal or off-diagonal, and $l$ labels the links of the lattice. Although these two terms are single-link operators, they may break the soft constraint when considering neighbors, so we have to regard the single-link operator as a multi-link operator instead with all closest neighbors as shown in Fig.~\ref{SF1}.

We can design the Monte Carlo algorithm to update vertices according to the soft constraint on the cells as shown in Supplementary Fig.~\ref{SF1}. 
Since the original sweeping cluster method always obeys the constraints without changing the number of dimers per site, adding such considerations for the terms in Eq.~(1) into the original sweeping cluster Monte Carlo method makes all samplings satisfy the soft constraint.


\noindent{\bf Stochastic analytic continuation.}
The main idea behind the stochastic analytic continuation (SAC) method~\cite{Sandvik1998a,Beach2004,Syljuasen2008,shao2022progress} is to obtain the optimal solution of the inverse Laplace transform via sampling dependent on importance of goodness. A set of imaginary-time correlation functions $G(\tau)$ can be obtained through the sweeping cluster QMC method first. The real-frequency spectral function and the imaginary-time correlation function are related by a  Laplace transformation as $G(\tau)=\int_0^\infty d\omega (e^{-\tau\omega}+e^{-(\beta-\tau)\omega}) S(\omega)/\pi$. We can inversely solve this equation by fitting a better spectral function. Assume the spectral function has a general form, $S(\omega)=\sum_{i} a_i \delta(\omega-\omega_i)$. We can obtain the optimal spectral function, \textit{i.e.}, the optimal choice of the set $\{a_i, \omega_i\}$ in the ansatz, numerically through sampling according to the importance of goodness of fit, with a simulated-annealing approach and with respect to the QMC errorbars of the imaginary-time correlation data $G(\tau)$. The reliability of such a QMC-SAC scheme has been extensively tested in various quantum many-body systems, such as the $1$D Heisenberg chain~\cite{Sandvik2015} compared to the Bethe ansatz, the $2$D Heisenberg model~\cite{Shao2017,Zhou2021amplitude} in comparison to exact diagonalization, field theoretical analysis and neutron scattering spectra in real square-lattice quantum magnets, deconfined quantum critical points~\cite{Shao2017,Ma2018a} and deconfined U($1$) spin liquid phases with emergent photon excitations~\cite{CJHuang2018}, $\mathbb{Z}_2$ quantum spin liquid models with fractionalized spectra~\cite{GYSun2018,YCWangVestigial2020,YCWang2021NC} via anyon condensation theory, and the quantum Ising model with direct comparison to neutron scattering and NMR experiments~\cite{Lih2020,ZHu2020}. We refer the readers to the technical descriptions available in the literature for the detailed documentation of our QMC+SAC scheme. 

\vspace{\baselineskip}
\noindent{\bf Data availability}

The data that support the findings of this study are available from the authors upon reasonable request.

\vspace{\baselineskip}
\noindent{\bf Code availability}

All numerical codes in this paper are available upon reasonable request to the authors.


\begin{thebibliography}{65}
\providecommand{\natexlab}[1]{#1}
\providecommand{\url}[1]{\texttt{#1}}
\expandafter\ifx\csname urlstyle\endcsname\relax
  \providecommand{\doi}[1]{doi: #1}\else
  \providecommand{\doi}{doi: \begingroup \urlstyle{rm}\Url}\fi

\bibitem[{Satzinger} et~al.(2021){Satzinger}, {Liu}, {Smith}, {Knapp},
  {Newman}, {Jones}, {Chen}, {Quintana}, {Mi}, {Dunsworth}, {Gidney},
  {Aleiner}, {Arute}, {Arya}, {Atalaya}, {Babbush}, {Bardin}, {Barends},
  {Basso}, {Bengtsson}, {Bilmes}, {Broughton}, {Buckley}, {Buell}, {Burkett},
  {Bushnell}, {Chiaro}, {Collins}, {Courtney}, {Demura}, {Derk}, {Eppens},
  {Erickson}, {Faoro}, {Farhi}, {Fowler}, {Foxen}, {Giustina}, {Greene},
  {Gross}, {Harrigan}, {Harrington}, {Hilton}, {Hong}, {Huang}, {Huggins},
  {Ioffe}, {Isakov}, {Jeffrey}, {Jiang}, {Kafri}, {Kechedzhi}, {Khattar},
  {Kim}, {Klimov}, {Korotkov}, {Kostritsa}, {Landhuis}, {Laptev}, {Locharla},
  {Lucero}, {Martin}, {McClean}, {McEwen}, {Miao}, {Mohseni}, {Montazeri},
  {Mruczkiewicz}, {Mutus}, {Naaman}, {Neeley}, {Neill}, {Niu},
  {O{\textquoteright}Brien}, {Opremcak}, {Pat{\'o}}, {Petukhov}, {Rubin},
  {Sank}, {Shvarts}, {Strain}, {Szalay}, {Villalonga}, {White}, {Yao}, {Yeh},
  {Yoo}, {Zalcman}, {Neven}, {Boixo}, {Megrant}, {Chen}, {Kelly},
  {Smelyanskiy}, {Kitaev}, {Knap}, {Pollmann}, and {Roushan}]{Roushan21}
K.~J. {Satzinger}, Y.~J. {Liu}, A.~{Smith}, C.~{Knapp}, M.~{Newman},
  C.~{Jones}, Z.~{Chen}, C.~{Quintana}, X.~{Mi}, A.~{Dunsworth}, C.~{Gidney},
  I.~{Aleiner}, F.~{Arute}, K.~{Arya}, J.~{Atalaya}, R.~{Babbush}, J.~C.
  {Bardin}, R.~{Barends}, J.~{Basso}, A.~{Bengtsson}, A.~{Bilmes},
  M.~{Broughton}, B.~B. {Buckley}, D.~A. {Buell}, B.~{Burkett}, N.~{Bushnell},
  B.~{Chiaro}, R.~{Collins}, W.~{Courtney}, S.~{Demura}, A.~R. {Derk},
  D.~{Eppens}, C.~{Erickson}, L.~{Faoro}, E.~{Farhi}, A.~G. {Fowler},
  B.~{Foxen}, M.~{Giustina}, A.~{Greene}, J.~A. {Gross}, M.~P. {Harrigan},
  S.~D. {Harrington}, J.~{Hilton}, S.~{Hong}, T.~{Huang}, W.~J. {Huggins},
  L.~B. {Ioffe}, S.~V. {Isakov}, E.~{Jeffrey}, Z.~{Jiang}, D.~{Kafri},
  K.~{Kechedzhi}, T.~{Khattar}, S.~{Kim}, P.~V. {Klimov}, A.~N. {Korotkov},
  F.~{Kostritsa}, D.~{Landhuis}, P.~{Laptev}, A.~{Locharla}, E.~{Lucero},
  O.~{Martin}, J.~R. {McClean}, M.~{McEwen}, K.~C. {Miao}, M.~{Mohseni},
  S.~{Montazeri}, W.~{Mruczkiewicz}, J.~{Mutus}, O.~{Naaman}, M.~{Neeley},
  C.~{Neill}, M.~Y. {Niu}, T.~E. {O{\textquoteright}Brien}, A.~{Opremcak},
  B.~{Pat{\'o}}, A.~{Petukhov}, N.~C. {Rubin}, D.~{Sank}, V.~{Shvarts},
  D.~{Strain}, M.~{Szalay}, B.~{Villalonga}, T.~C. {White}, Z.~{Yao}, P.~{Yeh},
  J.~{Yoo}, A.~{Zalcman}, H.~{Neven}, S.~{Boixo}, A.~{Megrant}, Y.~{Chen},
  J.~{Kelly}, V.~{Smelyanskiy}, A.~{Kitaev}, M.~{Knap}, F.~{Pollmann}, and
  P.~{Roushan}.
\newblock {Realizing topologically ordered states on a quantum processor}.
\newblock \emph{Science}, 374\penalty0 (6572):\penalty0 1237--1241, December
  2021.
\newblock \doi{10.1126/science.abi8378}.

\bibitem[{Semeghini} et~al.(2021){Semeghini}, {Levine}, {Keesling}, {Ebadi},
  {Wang}, {Bluvstein}, {Verresen}, {Pichler}, {Kalinowski}, {Samajdar},
  {Omran}, {Sachdev}, {Vishwanath}, {Greiner}, {Vuleti{\'c}}, and
  {Lukin}]{Semeghini21}
G.~{Semeghini}, H.~{Levine}, A.~{Keesling}, S.~{Ebadi}, T.~T. {Wang},
  D.~{Bluvstein}, R.~{Verresen}, H.~{Pichler}, M.~{Kalinowski}, R.~{Samajdar},
  A.~{Omran}, S.~{Sachdev}, A.~{Vishwanath}, M.~{Greiner}, V.~{Vuleti{\'c}},
  and M.~D. {Lukin}.
\newblock {Probing topological spin liquids on a programmable quantum
  simulator}.
\newblock \emph{Science}, 374\penalty0 (6572):\penalty0 1242--1247, December
  2021.
\newblock \doi{10.1126/science.abi8794}.

\bibitem[Read and Sachdev(1991)]{RS91}
N.~Read and Subir Sachdev.
\newblock {Large-$N$ expansion for frustrated quantum antiferromagnets}.
\newblock \emph{Phys. Rev. Lett.}, 66:\penalty0 1773--1776, Apr 1991.
\newblock \doi{10.1103/PhysRevLett.66.1773}.
\newblock URL \url{https://link.aps.org/doi/10.1103/PhysRevLett.66.1773}.

\bibitem[Wen(1991)]{wen1991}
X.~G. Wen.
\newblock Mean-field theory of spin-liquid states with finite energy gap and
  topological orders.
\newblock \emph{Phys. Rev. B}, 44:\penalty0 2664--2672, Aug 1991.
\newblock \doi{10.1103/PhysRevB.44.2664}.
\newblock URL \url{https://link.aps.org/doi/10.1103/PhysRevB.44.2664}.

\bibitem[Kitaev(2003)]{Kitaev1997}
Alexei Kitaev.
\newblock {Fault tolerant quantum computation by anyons}.
\newblock \emph{Ann. Phys.}, 303:\penalty0 2--30, 2003.
\newblock \doi{10.1016/S0003-4916(02)00018-0}.

\bibitem[Jalabert and Sachdev(1991)]{RJSS91}
Rodolfo~A. Jalabert and Subir Sachdev.
\newblock {Spontaneous alignment of frustrated bonds in an anisotropic,
  three-dimensional Ising model}.
\newblock \emph{Phys. Rev. B}, 44:\penalty0 686--690, Jul 1991.
\newblock \doi{10.1103/PhysRevB.44.686}.
\newblock URL \url{http://link.aps.org/doi/10.1103/PhysRevB.44.686}.

\bibitem[{Sachdev} and {Vojta}(2000)]{MVSS99}
S.~{Sachdev} and M.~{Vojta}.
\newblock {Translational symmetry breaking in two-dimensional antiferromagnets
  and superconductors}.
\newblock \emph{J. Phys. Soc. Jpn.}, 69, Supp. B:\penalty0 1, 2000.
\newblock URL \url{https://arxiv.org/abs/cond-mat/9910231}.

\bibitem[{Senthil} and {Fisher}(2000)]{TSMPAF99}
T.~{Senthil} and M.~P.~A. {Fisher}.
\newblock {$\mathbb{Z}_{2}$ gauge theory of electron fractionalization in
  strongly correlated systems}.
\newblock \emph{Phys. Rev. B}, 62:\penalty0 7850, September 2000.
\newblock \doi{10.1103/PhysRevB.62.7850}.

\bibitem[{Moessner} et~al.(2001){Moessner}, {Sondhi}, and
  {Fradkin}]{Moessner01}
R.~{Moessner}, S.~L. {Sondhi}, and Eduardo {Fradkin}.
\newblock {Short-ranged resonating valence bond physics, quantum dimer models,
  and Ising gauge theories}.
\newblock \emph{Phys. Rev. B}, 65\penalty0 (2):\penalty0 024504, December 2001.
\newblock \doi{10.1103/PhysRevB.65.024504}.

\bibitem[Essin and Hermele(2014)]{Essin2014}
Andrew~M. Essin and Michael Hermele.
\newblock Spectroscopic signatures of crystal momentum fractionalization.
\newblock \emph{Phys. Rev. B}, 90:\penalty0 121102, Sep 2014.
\newblock \doi{10.1103/PhysRevB.90.121102}.
\newblock URL \url{https://link.aps.org/doi/10.1103/PhysRevB.90.121102}.

\bibitem[{Mei} and {Wen}(2015)]{JWMei2015}
Jia-Wei {Mei} and Xiao-Gang {Wen}.
\newblock {Fractionalized spin-wave continuum in spin liquid states on the
  kagome lattice}.
\newblock \emph{arXiv e-prints}, art. arXiv:1507.03007, July 2015.

\bibitem[Sun et~al.(2018)Sun, Wang, Fang, Qi, Cheng, and Meng]{GYSun2018}
Guang-Yu Sun, Yan-Cheng Wang, Chen Fang, Yang Qi, Meng Cheng, and Zi~Yang Meng.
\newblock {Dynamical Signature of Symmetry Fractionalization in Frustrated
  Magnets}.
\newblock \emph{Phys. Rev. Lett.}, 121:\penalty0 077201, Aug 2018.
\newblock \doi{10.1103/PhysRevLett.121.077201}.
\newblock URL \url{https://link.aps.org/doi/10.1103/PhysRevLett.121.077201}.

\bibitem[{Wang} et~al.(2021){Wang}, {Cheng}, {Witczak-Krempa}, and
  {Meng}]{YCWang2021NC}
Yan-Cheng {Wang}, Meng {Cheng}, William {Witczak-Krempa}, and Zi~Yang {Meng}.
\newblock {Fractionalized conductivity and emergent self-duality near
  topological phase transitions}.
\newblock \emph{Nat. Commun.}, 12:\penalty0 5347, 2021.
\newblock \doi{10.1038/s41467-021-25707-z}.
\newblock URL \url{https://doi.org/10.1038/s41467-021-25707-z}.

\bibitem[{Wang} et~al.(2017){Wang}, {Fang}, {Cheng}, {Qi}, and
  {Meng}]{YCWang2017QSL}
Yan-Cheng {Wang}, Chen {Fang}, Meng {Cheng}, Yang {Qi}, and Zi~Yang {Meng}.
\newblock {Topological Spin Liquid with Symmetry-Protected Edge States}.
\newblock \emph{arXiv e-prints}, art. arXiv:1701.01552, Jan 2017.

\bibitem[Wang et~al.(2018)Wang, Zhang, Pollmann, Cheng, and Meng]{YCWang2018}
Yan-Cheng Wang, Xue-Feng Zhang, Frank Pollmann, Meng Cheng, and Zi~Yang Meng.
\newblock {Quantum Spin Liquid with Even Ising Gauge Field Structure on Kagome
  Lattice}.
\newblock \emph{Phys. Rev. Lett.}, 121:\penalty0 057202, Aug 2018.
\newblock \doi{10.1103/PhysRevLett.121.057202}.
\newblock URL \url{https://link.aps.org/doi/10.1103/PhysRevLett.121.057202}.

\bibitem[Essin and Hermele(2013)]{Essin:2013rca}
Andrew~M. Essin and Michael Hermele.
\newblock {Classifying fractionalization: Symmetry classification of gapped
  $\mathbb{Z}_2$ spin liquids in two dimensions}.
\newblock \emph{Phys. Rev. B}, 87\penalty0 (10):\penalty0 104406, 2013.
\newblock \doi{10.1103/PhysRevB.87.104406}.

\bibitem[Zaletel and Vishwanath(2015)]{Zaletel:2014epa}
Michael~P. Zaletel and Ashvin Vishwanath.
\newblock {Constraints on topological order in Mott Insulators}.
\newblock \emph{Phys. Rev. Lett.}, 114\penalty0 (7):\penalty0 077201, 2015.
\newblock \doi{10.1103/PhysRevLett.114.077201}.

\bibitem[Cheng et~al.(2016)Cheng, Zaletel, Barkeshli, Vishwanath, and
  Bonderson]{Cheng:2015kce}
Meng Cheng, Michael Zaletel, Maissam Barkeshli, Ashvin Vishwanath, and Parsa
  Bonderson.
\newblock {Translational Symmetry and Microscopic Constraints on
  Symmetry-Enriched Topological Phases: A View from the Surface}.
\newblock \emph{Phys. Rev. X}, 6\penalty0 (4):\penalty0 041068, 2016.
\newblock \doi{10.1103/PhysRevX.6.041068}.

\bibitem[Qi and Cheng(2018)]{QiMeng16}
Yang Qi and Meng Cheng.
\newblock {Classification of symmetry fractionalization in gapped
  ${\mathbb{Z}}_{2}$ spin liquids}.
\newblock \emph{Phys. Rev. B}, 97:\penalty0 115138, Mar 2018.
\newblock \doi{10.1103/PhysRevB.97.115138}.
\newblock URL \url{https://link.aps.org/doi/10.1103/PhysRevB.97.115138}.

\bibitem[Bulmash and Barkeshli(2020)]{Bulmash:2020flp}
Daniel Bulmash and Maissam Barkeshli.
\newblock {Absolute anomalies in (2+1)D symmetry-enriched topological states
  and exact (3+1)D constructions}.
\newblock \emph{Phys. Rev. Research}, 2\penalty0 (4):\penalty0 043033, 2020.
\newblock \doi{10.1103/PhysRevResearch.2.043033}.

\bibitem[Rokhsar and Kivelson(1988)]{RK88}
Daniel~S. Rokhsar and Steven~A. Kivelson.
\newblock {Superconductivity and the Quantum Hard-Core Dimer Gas}.
\newblock \emph{Phys. Rev. Lett.}, 61:\penalty0 2376--2379, Nov 1988.
\newblock \doi{10.1103/PhysRevLett.61.2376}.
\newblock URL \url{http://link.aps.org/doi/10.1103/PhysRevLett.61.2376}.

\bibitem[Moessner and Raman(2011)]{moessner2011quantum}
Roderich Moessner and Kumar~S Raman.
\newblock Quantum dimer models.
\newblock In \emph{{Introduction to Frustrated Magnetism}}, pages 437--479.
  Springer, 2011.
\newblock \doi{10.1007/978-3-642-10589-0_17}.

\bibitem[{Moessner} and {Sondhi}(2001)]{RMSLS01}
R.~{Moessner} and S.~L. {Sondhi}.
\newblock {Resonating Valence Bond Phase in the Triangular Lattice Quantum
  Dimer Model}.
\newblock \emph{Phys. Rev. Lett.}, 86:\penalty0 1881, February 2001.
\newblock \doi{10.1103/PhysRevLett.86.1881}.

\bibitem[Roychowdhury et~al.(2015)Roychowdhury, Bhattacharjee, and
  Pollmann]{roychowdhury2015z}
Krishanu Roychowdhury, Subhro Bhattacharjee, and Frank Pollmann.
\newblock {${\mathbb{Z}}_{2}$ topological liquid of hard-core bosons on a
  kagome lattice at $1/3$ filling}.
\newblock \emph{Phys. Rev. B}, 92:\penalty0 075141, Aug 2015.
\newblock \doi{10.1103/PhysRevB.92.075141}.
\newblock URL \url{https://link.aps.org/doi/10.1103/PhysRevB.92.075141}.

\bibitem[Plat et~al.(2015)Plat, Alet, Capponi, and Totsuka]{Plat2015z2}
X.~Plat, F.~Alet, S.~Capponi, and K.~Totsuka.
\newblock Magnetization plateaus of an easy-axis kagome antiferromagnet with
  extended interactions.
\newblock \emph{Phys. Rev. B}, 92:\penalty0 174402, Nov 2015.
\newblock \doi{10.1103/PhysRevB.92.174402}.
\newblock URL \url{https://link.aps.org/doi/10.1103/PhysRevB.92.174402}.

\bibitem[{Fendley} et~al.(2004){Fendley}, {Sengupta}, and
  {Sachdev}]{fendley2004competing}
P.~{Fendley}, K.~{Sengupta}, and S.~{Sachdev}.
\newblock {Competing density-wave orders in a one-dimensional hard-boson
  model}.
\newblock \emph{Phys. Rev. B}, 69\penalty0 (7):\penalty0 075106, February 2004.
\newblock \doi{10.1103/PhysRevB.69.075106}.

\bibitem[{Bernien} et~al.(2017){Bernien}, {Schwartz}, {Keesling}, {Levine},
  {Omran}, {Pichler}, {Choi}, {Zibrov}, {Endres}, {Greiner}, {Vuleti{\'c}}, and
  {Lukin}]{Bernien17}
Hannes {Bernien}, Sylvain {Schwartz}, Alexander {Keesling}, Harry {Levine},
  Ahmed {Omran}, Hannes {Pichler}, Soonwon {Choi}, Alexander~S. {Zibrov},
  Manuel {Endres}, Markus {Greiner}, Vladan {Vuleti{\'c}}, and Mikhail~D.
  {Lukin}.
\newblock {Probing many-body dynamics on a 51-atom quantum simulator}.
\newblock \emph{Nature}, 551\penalty0 (7682):\penalty0 579--584, November 2017.
\newblock \doi{10.1038/nature24622}.

\bibitem[Samajdar et~al.(2021)Samajdar, Ho, Pichler, Lukin, and
  Sachdev]{Samajdar:2020hsw}
Rhine Samajdar, Wen~Wei Ho, Hannes Pichler, Mikhail~D. Lukin, and Subir
  Sachdev.
\newblock {Quantum phases of Rydberg atoms on a kagome lattice}.
\newblock \emph{Proc. Natl. Acad. Sci. U.S.A.}, 118:\penalty0 e2015785118,
  2021.
\newblock \doi{10.1073/pnas.2015785118}.

\bibitem[Verresen et~al.(2021)Verresen, Lukin, and
  Vishwanath]{Verresen:2020dmk}
Ruben Verresen, Mikhail~D. Lukin, and Ashvin Vishwanath.
\newblock {Prediction of Toric Code Topological Order from Rydberg Blockade}.
\newblock \emph{Phys. Rev. X}, 11\penalty0 (3):\penalty0 031005, 2021.
\newblock \doi{10.1103/PhysRevX.11.031005}.

\bibitem[Samajdar et~al.(2022)Samajdar, Joshi, Teng, and Sachdev]{Z2GT}
Rhine Samajdar, Darshan~G. Joshi, Yanting Teng, and Subir Sachdev.
\newblock {Emergent $\mathbb{Z}_2$ gauge theories and topological excitations
  in Rydberg atom arrays}.
\newblock \emph{arXiv:2204.00632 [cond-mat.quant-gas]}, 2022.
\newblock URL \url{https://arxiv.org/abs/2204.00632}.


\bibitem[Ralko et~al.(2005)Ralko, Ferrero, Becca, Ivanov, and Mila]{Ralko2005}
Arnaud Ralko, Michel Ferrero, Federico Becca, Dmitri Ivanov, and Fr\'ed\'eric
  Mila.
\newblock Zero-temperature properties of the quantum dimer model on the
  triangular lattice.
\newblock \emph{Phys. Rev. B}, 71:\penalty0 224109, Jun 2005.
\newblock \doi{10.1103/PhysRevB.71.224109}.
\newblock URL \url{https://link.aps.org/doi/10.1103/PhysRevB.71.224109}.

\bibitem[{Papanikolaou} et~al.(2014){Papanikolaou}, {Charrier}, and
  {Fradkin}]{fradkin14}
Stefanos {Papanikolaou}, Daniel {Charrier}, and Eduardo {Fradkin}.
\newblock {Ising nematic fluid phase of hard-core dimers on the square
  lattice}.
\newblock \emph{Phys. Rev. B}, 89\penalty0 (3):\penalty0 035128, January 2014.
\newblock \doi{10.1103/PhysRevB.89.035128}.

\bibitem[Moessner and Sondhi(2001{\natexlab{a}})]{MoessnerSondhi2001a}
R.~Moessner and S.~L. Sondhi.
\newblock Resonating valence bond phase in the triangular lattice quantum dimer
  model.
\newblock \emph{Phys. Rev. Lett.}, 86:\penalty0 1881--1884, Feb
  2001{\natexlab{a}}.
\newblock \doi{10.1103/PhysRevLett.86.1881}.
\newblock URL \url{https://link.aps.org/doi/10.1103/PhysRevLett.86.1881}.

\bibitem[Moessner and Sondhi(2001{\natexlab{b}})]{MoessnerSondhi2001b}
R.~Moessner and S.~L. Sondhi.
\newblock Ising models of quantum frustration.
\newblock \emph{Phys. Rev. B}, 63:\penalty0 224401, May 2001{\natexlab{b}}.
\newblock \doi{10.1103/PhysRevB.63.224401}.
\newblock URL \url{https://link.aps.org/doi/10.1103/PhysRevB.63.224401}.

\bibitem[Ralko et~al.(2006)Ralko, Ferrero, Becca, Ivanov, and Mila]{Ralko2006}
Arnaud Ralko, Michel Ferrero, Federico Becca, Dmitri Ivanov, and Fr\'ed\'eric
  Mila.
\newblock Dynamics of the quantum dimer model on the triangular lattice: Soft
  modes and local resonating valence-bond correlations.
\newblock \emph{Phys. Rev. B}, 74:\penalty0 134301, Oct 2006.
\newblock \doi{10.1103/PhysRevB.74.134301}.
\newblock URL \url{https://link.aps.org/doi/10.1103/PhysRevB.74.134301}.

\bibitem[Ralko et~al.(2008)Ralko, Poilblanc, and Moessner]{Ralko2008}
A.~Ralko, D.~Poilblanc, and R.~Moessner.
\newblock {Generic Mixed Columnar-Plaquette Phases in Rokhsar-Kivelson Models}.
\newblock \emph{Phys. Rev. Lett.}, 100:\penalty0 037201, Jan 2008.
\newblock \doi{10.1103/PhysRevLett.100.037201}.
\newblock URL \url{https://link.aps.org/doi/10.1103/PhysRevLett.100.037201}.

\bibitem[Yan et~al.(2021{\natexlab{a}})Yan, Wang, Ma, Qi, and Meng]{ZY2020}
Zheng Yan, Yan-Cheng Wang, Nvsen Ma, Yang Qi, and Zi~Yang Meng.
\newblock Topological phase transition and single/multi anyon dynamics of
  ${Z}_2$ spin liquid.
\newblock \emph{npj Quantum Mater.}, page~39, 2021{\natexlab{a}}.
\newblock \doi{10.1038/s41535-021-00338-1}.
\newblock URL \url{https://doi.org/10.1038/s41535-021-00338-1}.

\bibitem[Yan et~al.(2019)Yan, Wu, Liu, Sylju\aa{}sen, Lou, and
  Chen]{ZY2019sweeping}
Zheng Yan, Yongzheng Wu, Chenrong Liu, Olav~F. Sylju\aa{}sen, Jie Lou, and Yan
  Chen.
\newblock Sweeping cluster algorithm for quantum spin systems with strong
  geometric restrictions.
\newblock \emph{Phys. Rev. B}, 99:\penalty0 165135, Apr 2019.
\newblock \doi{10.1103/PhysRevB.99.165135}.
\newblock URL \url{https://link.aps.org/doi/10.1103/PhysRevB.99.165135}.

\bibitem[Yan(2022)]{ZY2020improved}
Zheng Yan.
\newblock Global scheme of sweeping cluster algorithm to sample among
  topological sectors.
\newblock \emph{Phys. Rev. B}, 105:\penalty0 184432, May 2022.
\newblock \doi{10.1103/PhysRevB.105.184432}.
\newblock URL \url{https://link.aps.org/doi/10.1103/PhysRevB.105.184432}.

\bibitem[Yan et~al.(2021{\natexlab{b}})Yan, Zhou, Sylju\aa{}sen, Zhang, Yuan,
  Lou, and Chen]{ZY2021mixed}
Zheng Yan, Zheng Zhou, Olav~F. Sylju\aa{}sen, Junhao Zhang, Tianzhong Yuan, Jie
  Lou, and Yan Chen.
\newblock Widely existing mixed phase structure of the quantum dimer model on a
  square lattice.
\newblock \emph{Phys. Rev. B}, 103:\penalty0 094421, Mar 2021{\natexlab{b}}.
\newblock \doi{10.1103/PhysRevB.103.094421}.
\newblock URL \url{https://link.aps.org/doi/10.1103/PhysRevB.103.094421}.


\bibitem[Beach(2004)]{Beach2004}
K.~S.~D. Beach.
\newblock Identifying the maximum entropy method as a special limit of
  stochastic analytic continuation.
\newblock \emph{arXiv:cond-mat/0403055 [cond-mat.str-el]}, 2004.
\newblock URL \url{https://arxiv.org/abs/cond-mat/0403055}.

\bibitem[Shao et~al.(2017{\natexlab{a}})Shao, Qin, Capponi, Chesi, Meng, and
  Sandvik]{Shao2017nearly}
Hui Shao, Yan~Qi Qin, Sylvain Capponi, Stefano Chesi, Zi~Yang Meng, and
  Anders~W. Sandvik.
\newblock Nearly deconfined spinon excitations in the square-lattice spin-$1/2$
  heisenberg antiferromagnet.
\newblock \emph{Phys. Rev. X}, 7:\penalty0 041072, Dec 2017{\natexlab{a}}.
\newblock \doi{10.1103/PhysRevX.7.041072}.
\newblock URL \url{https://link.aps.org/doi/10.1103/PhysRevX.7.041072}.

\bibitem[Zhou et~al.(2021{\natexlab{a}})Zhou, Yan, Wu, Sun, Starykh, and
  Meng]{Zhou2021amplitude}
Chengkang Zhou, Zheng Yan, Han-Qing Wu, Kai Sun, Oleg~A. Starykh, and Zi~Yang
  Meng.
\newblock Amplitude mode in quantum magnets via dimensional crossover.
\newblock \emph{Phys. Rev. Lett.}, 126:\penalty0 227201, Jun
  2021{\natexlab{a}}.
\newblock \doi{10.1103/PhysRevLett.126.227201}.
\newblock URL \url{https://link.aps.org/doi/10.1103/PhysRevLett.126.227201}.

\bibitem[Wang et~al.(2021{\natexlab{a}})Wang, Yan, Wang, Qi, and
  Meng]{YCWang2021vestigial}
Yan-Cheng Wang, Zheng Yan, Chenjie Wang, Yang Qi, and Zi~Yang Meng.
\newblock Vestigial anyon condensation in kagome quantum spin liquids.
\newblock \emph{Phys. Rev. B}, 103:\penalty0 014408, Jan 2021{\natexlab{a}}.
\newblock \doi{10.1103/PhysRevB.103.014408}.
\newblock URL \url{https://link.aps.org/doi/10.1103/PhysRevB.103.014408}.

\bibitem[Shao and Sandvik(2022)]{shao2022progress}
Hui Shao and Anders~W Sandvik.
\newblock Progress on stochastic analytic continuation of quantum Monte Carlo
  data.
\newblock \emph{arXiv preprint arXiv:2202.09870}, 2022.

\bibitem[Ivanov(2004)]{Ivanov2004PRB}
D.~A. Ivanov.
\newblock Vortexlike elementary excitations in the Rokhsar-Kivelson dimer model on the triangular lattice.
\newblock \emph{Phys. Rev. B}, 70:\penalty0 094430, Sep 2004.
\newblock \doi{10.1103/PhysRevB.70.094430}.
\newblock URL \url{https://link.aps.org/doi/10.1103/PhysRevB.70.094430}.

\bibitem[Ralko et~al.(2007)Ralko, Ferrero, Becca, Ivanov, and
  Mila]{Ralko2007PRB}
Arnaud Ralko, Michel Ferrero, Federico Becca, Dmitri Ivanov, and Fr\'ed\'eric
  Mila.
\newblock Crystallization of the resonating valence bond liquid as vortex
  condensation.
\newblock \emph{Phys. Rev. B}, 76:\penalty0 140404, Oct 2007.
\newblock \doi{10.1103/PhysRevB.76.140404}.
\newblock URL \url{https://link.aps.org/doi/10.1103/PhysRevB.76.140404}.

\bibitem[Yan et~al.(2022{\natexlab{a}})Yan, Meng, Huse, and Chan]{ZY2022hQDM}
Zheng Yan, Zi~Yang Meng, David~A. Huse, and Amos Chan.
\newblock Height-conserving quantum dimer models.
\newblock \emph{Phys. Rev. B}, 106:\penalty0 L041115, Jul 2022{\natexlab{a}}.
\newblock \doi{10.1103/PhysRevB.106.L041115}.
\newblock URL \url{https://link.aps.org/doi/10.1103/PhysRevB.106.L041115}.

\bibitem[Yan et~al.(2022{\natexlab{b}})Yan, Ran, Wang, Samajdar, Rong, Sachdev,
  Qi, and Meng]{ZY2022loop}
Zheng Yan, Xiaoxue Ran, Yan-Cheng Wang, Rhine Samajdar, Junchen Rong, Subir
  Sachdev, Yang Qi, and Zi~Yang Meng.
\newblock Fully packed quantum loop model on the triangular lattice: Hidden
  vison plaquette phase and cubic phase transitions.
\newblock \emph{arXiv preprint arXiv:2205.04472}, 2022{\natexlab{b}}.

\bibitem[Zhou et~al.(2022)Zhou, Liu, Yan, Chen, and
  Zhang]{zhou2020quantumstring}
Zheng Zhou, Changle Liu, Zheng Yan, Yan Chen, and Xue-Feng Zhang.
\newblock Quantum dynamics of topological strings in a frustrated Ising
  antiferromagnet.
\newblock \emph{npj Quantum Mater.}, 7\penalty0 (1):\penalty0 1--7, 2022.
\newblock URL \url{https://www.nature.com/articles/s41535-022-00465-3}.

\bibitem[Zhou et~al.(2020)Zhou, Liu, Yan, Chen, and Zhang]{zhou2020quantum}
Zheng Zhou, Dong-Xu Liu, Zheng Yan, Yan Chen, and Xue-Feng Zhang.
\newblock Quantum tricriticality of incommensurate phase induced by quantum
  domain walls in frustrated Ising magnetism.
\newblock \emph{arXiv preprint arXiv:2005.11133}, 2020.

\bibitem[Yan et~al.(2021{\natexlab{c}})Yan, Zhou, Wang, Meng, and
  Zhang]{yan2022targeting}
Zheng Yan, Zheng Zhou, Yan-Cheng Wang, Zi~Yang Meng, and Xue-Feng Zhang.
\newblock Sweeping quantum annealing algorithm for constrained optimization
  problems.
\newblock \emph{arXiv preprint arXiv:2105.07134}, 2021{\natexlab{c}}.

\bibitem[Zhou et~al.(2021{\natexlab{b}})Zhou, Yan, Liu, Chen, and
  Zhang]{zhou2021emergent}
Zheng Zhou, Zheng Yan, Changle Liu, Yan Chen, and Xue-Feng Zhang.
\newblock Emergent Rokhsar-Kivelson point in realistic quantum ising models.
\newblock \emph{arXiv preprint arXiv:2106.05518}, 2021{\natexlab{b}}.

\bibitem[Sandvik(2010)]{Sandvik2010b}
A.~W. Sandvik.
\newblock Computational studies of quantum spin systems.
\newblock \emph{AIP Conf. Proc.}, 1297\penalty0 (1):\penalty0 135--338, 2010.
\newblock \doi{10.1063/1.3518900}.

\bibitem[Sandvik(1998{\natexlab{b}})]{Sandvik1998a}
Anders~W. Sandvik.
\newblock {Stochastic method for analytic continuation of quantum Monte Carlo
  data}.
\newblock \emph{Phys. Rev. B}, 57:\penalty0 10287--10290, May
  1998{\natexlab{b}}.
\newblock \doi{10.1103/PhysRevB.57.10287}.
\newblock URL \url{http://link.aps.org/doi/10.1103/PhysRevB.57.10287}.

\bibitem[Sylju\aa{}sen(2008)]{Syljuasen2008}
Olav~F. Sylju\aa{}sen.
\newblock Using the average spectrum method to extract dynamics from quantum
  monte carlo simulations.
\newblock \emph{Phys. Rev. B}, 78:\penalty0 174429, Nov 2008.
\newblock \doi{10.1103/PhysRevB.78.174429}.
\newblock URL \url{https://link.aps.org/doi/10.1103/PhysRevB.78.174429}.

\bibitem[Sandvik(2016)]{Sandvik2015}
Anders~W. Sandvik.
\newblock Constrained sampling method for analytic continuation.
\newblock \emph{Phys. Rev. E}, 94:\penalty0 063308, Dec 2016.
\newblock \doi{10.1103/PhysRevE.94.063308}.
\newblock URL \url{https://link.aps.org/doi/10.1103/PhysRevE.94.063308}.

\bibitem[Shao et~al.(2017{\natexlab{b}})Shao, Qin, Capponi, Chesi, Meng, and
  Sandvik]{Shao2017}
Hui Shao, Yan~Qi Qin, Sylvain Capponi, Stefano Chesi, Zi~Yang Meng, and
  Anders~W. Sandvik.
\newblock Nearly deconfined spinon excitations in the square-lattice spin-$1/2$
  Heisenberg antiferromagnet.
\newblock \emph{Phys. Rev. X}, 7:\penalty0 041072, Dec 2017{\natexlab{b}}.
\newblock \doi{10.1103/PhysRevX.7.041072}.
\newblock URL \url{https://link.aps.org/doi/10.1103/PhysRevX.7.041072}.

\bibitem[Ma et~al.(2018)Ma, Sun, You, Xu, Vishwanath, Sandvik, and
  Meng]{Ma2018a}
Nvsen Ma, Guang-Yu Sun, Yi-Zhuang You, Cenke Xu, Ashvin Vishwanath, Anders~W.
  Sandvik, and Zi~Yang Meng.
\newblock Dynamical signature of fractionalization at a deconfined quantum
  critical point.
\newblock \emph{Phys. Rev. B}, 98:\penalty0 174421, Nov 2018.
\newblock \doi{10.1103/PhysRevB.98.174421}.
\newblock URL \url{https://link.aps.org/doi/10.1103/PhysRevB.98.174421}.

\bibitem[Huang et~al.(2018)Huang, Deng, Wan, and Meng]{CJHuang2018}
Chun-Jiong Huang, Youjin Deng, Yuan Wan, and Zi~Yang Meng.
\newblock {Dynamics of Topological Excitations in a Model Quantum Spin Ice}.
\newblock \emph{Phys. Rev. Lett.}, 120:\penalty0 167202, Apr 2018.
\newblock \doi{10.1103/PhysRevLett.120.167202}.
\newblock URL \url{https://link.aps.org/doi/10.1103/PhysRevLett.120.167202}.

\bibitem[Wang et~al.(2021{\natexlab{b}})Wang, Yan, Wang, Qi, and
  Meng]{YCWangVestigial2020}
Yan-Cheng Wang, Zheng Yan, Chenjie Wang, Yang Qi, and Zi~Yang Meng.
\newblock Vestigial anyon condensation in kagome quantum spin liquids.
\newblock \emph{Phys. Rev. B}, 103:\penalty0 014408, Jan 2021{\natexlab{b}}.
\newblock \doi{10.1103/PhysRevB.103.014408}.
\newblock URL \url{https://link.aps.org/doi/10.1103/PhysRevB.103.014408}.

\bibitem[Li et~al.(2020)Li, Liao, Chen, Zeng, Sheng, Qi, Meng, and Li]{Lih2020}
Han Li, Yuan~Da Liao, Bin-Bin Chen, Xu-Tao Zeng, Xian-Lei Sheng, Yang Qi,
  Zi~Yang Meng, and Wei Li.
\newblock {Kosterlitz-Thouless} melting of magnetic order in the triangular
  quantum {Ising} material {TmMgGaO$_4$}.
\newblock \emph{Nat. Commun.}, 11\penalty0 (1):\penalty0 1111, 2020.
\newblock \doi{10.1038/s41467-020-14907-8}.
\newblock URL \url{https://doi.org/10.1038/s41467-020-14907-8}.

\bibitem[Hu et~al.(2020)Hu, Ma, Liao, Li, Ma, Cui, Shangguan, Huang, Qi, Li,
  Meng, Wen, and Yu]{ZHu2020}
Ze~Hu, Zhen Ma, Yuan-Da Liao, Han Li, Chunsheng Ma, Yi~Cui, Yanyan Shangguan,
  Zhentao Huang, Yang Qi, Wei Li, Zi~Yang Meng, Jinsheng Wen, and Weiqiang Yu.
\newblock {Evidence of the Berezinskii-Kosterlitz-Thouless phase in a
  frustrated magnet}.
\newblock \emph{Nat. Commun.}, 11:\penalty0 5631, 2020.
\newblock \doi{10.1038/s41467-020-19380-x}.
\newblock URL \url{https://doi.org/10.1038/s41467-020-19380-x}.

\end{thebibliography}

\vspace{\baselineskip}
\noindent{\bf Acknowledgements}

R.S. and S.S. are supported by the U.S. Department of Energy under Grant DE-SC0019030 and thank their coauthors in earlier collaborations \cite{Samajdar:2020hsw,Semeghini21}. Z.Y. and Z.Y.M. acknowledge support from the Research Grants
Counci of Hong Kong SAR of China (Grant Nos.~17303019, 17301420, 17301721 and AoE/P-701/20), the K. C. Wong Education Foundation (Grant No.~GJTD-2020-01) and the Seed Funding ``Quantum-Inspired explainable-AI" at the HKU-TCL Joint Research Centre for Artificial Intelligence. Y.C.W. acknowledges the supports from the NSFC under Grant Nos.~11804383 and 11975024. Y.C.W. and Z.Y. thank the support of Beihang Hangzhou Innovation Institute Yuhang. We thank Beijng PARATERA Tech CO.,Ltd., the supercomputing system in the High-performance Computing Centre of Beihang Hangzhou Innovation Institute Yuhang, the HPC2021 system under the Information Technology Services at the University of Hong Kong and the Tianhe-II platform at the National Supercomputer Center in Guangzhou for their technical support and generous allocation of CPU time.

\vspace{\baselineskip}
\noindent{\bf Author Contributions}

R.S., S.S. and Z.Y.M. initiated the work. Z.Y. developed the QMC algorithm for soft constraint. Z.Y. and Y.C.W. performed the computational simulations. All authors contributed to the analysis of the results. S.S. and Z.Y.M. supervised the project.

\vspace{\baselineskip}
\noindent{\bf Competing interests}

The authors declare no competing interests.\\


\end{document}


\vspace{\baselineskip}
\title{Supplementary Information for\\
``Triangular lattice quantum dimer model with variable dimer density''}

\author{Zheng Yan}
\affiliation{Department of Physics and HKU-UCAS Joint Institute of Theoretical and Computational Physics, The University of Hong Kong, Pokfulam Road, Hong Kong SAR, China}

\author{Rhine Samajdar}
\affiliation{Department of Physics, Harvard University, Cambridge MA 02138, USA}

\author{Yan-Cheng Wang}
\affiliation{Beihang Hangzhou Innovation Institute Yuhang, Hangzhou 310023, China}

\author{Subir Sachdev}
\email{sachdev@g.harvard.edu}
\affiliation{Department of Physics, Harvard University, Cambridge MA 02138, USA}
\affiliation{School of Natural Sciences, Institute for Advanced Study, Princeton, NJ 08540, USA}

\author{Zi Yang Meng}
\email{zymeng@hku.hk}
\affiliation{Department of Physics and HKU-UCAS Joint Institute of Theoretical and Computational Physics, The University of Hong Kong, Pokfulam Road, Hong Kong SAR, China}
\maketitle


\vspace{\baselineskip}
\noindent {\bf Supplementary Note 1: Relation between the dimer spectra and vison-pair correlations.}

In this section, we examine two dynamical correlation functions. The first one is the conventional dimer correlation: defining the dimer operator as $D_i=1$ ($=0$) when there is a (no) dimer on the link $i$, the dimer correlation function is given by $C_d(r_{i,j},\tau) =\sum_{i,j} \langle D_i(\tau)D_j(0)\rangle-\langle D_i\rangle^2$. Then, $C_d(\mathbf{q},\tau)$ can be computed via a Fourier transformation, following which the excitation spectrum $C_d(\mathbf{q},\omega)$ is obtained using stochastic analytic continuation (SAC).

The second quantity of interest is the correlation function of another differently  defined ``dimer", namely, the vison-pair correlation function. This ``dimer" is the vison-convolution (VC) operator, which is defined as $D^{vc}_{i}=V_{i_1}V_{i_2}d_{i}$, where $d_{i}$\,$=$\,$\pm 1$ when there is no/one dimer on link $i$ of the reference configuration (for the one-dimer-per-site case, we have to choose a reference configuration to fix the gauge; for the two-dimer-per-site case, $d_{i}=1\, \forall\, i$). The idea is that if two visons are close to each other, sharing the same link, then $D^{vc}_i$ on the link $i$ can be represented as the product of these two vison operators, with $i_1$ and $i_2$ being the triangular plaquettes closest to the link $i$.  Assuming the interaction of the visons is weak, this correlation function $C^{vc}_d(r_{i,j},\tau)=\langle D^{vc}_i(0)D^{vc}_j(\tau)\rangle-\langle D^{vc}_i(0)\rangle^2=\langle V_{i_1}(0)V_{i_2}(0)d_iV_{j_1}(\tau)V_{j_2}(\tau)d_j\rangle-\langle V_{i_1}(0)V_{i_2}(0)\rangle^2$ can be computed using Wick's theorem as the convolution of two vison operators,
\begin{equation}
C^{vc}_d(r_{i,j},\tau)=\langle V_{i_1}(0)V_{j_1}(\tau)\rangle \langle V_{i_2}(0)V_{j_2}(\tau)\rangle d_id_j +\langle V_{i_1}(0)V_{j_2}(\tau)\rangle \langle V_{i_2}(0)V_{j_1}(\tau)\rangle d_id_j.
\label{eq:eq2}
\end{equation}
Here, $d_i$ is constant for link $i$ under the choice of a gauge, and can be taken outside the brackets. 

Comparing the two abovementioned correlation functions, we see that one dimer can thus be treated as a vison pair deep in the quantum spin liquid (QSL) phase if the interaction effects among the visons are sufficiently weak. Further details in this regard can be found in the discussion accompanying Eq.~(2) of  Supplementary Ref.~\onlinecite{ZY2020}.

\begin{figure*}[t]
\centering
\includegraphics[width=0.5\columnwidth]{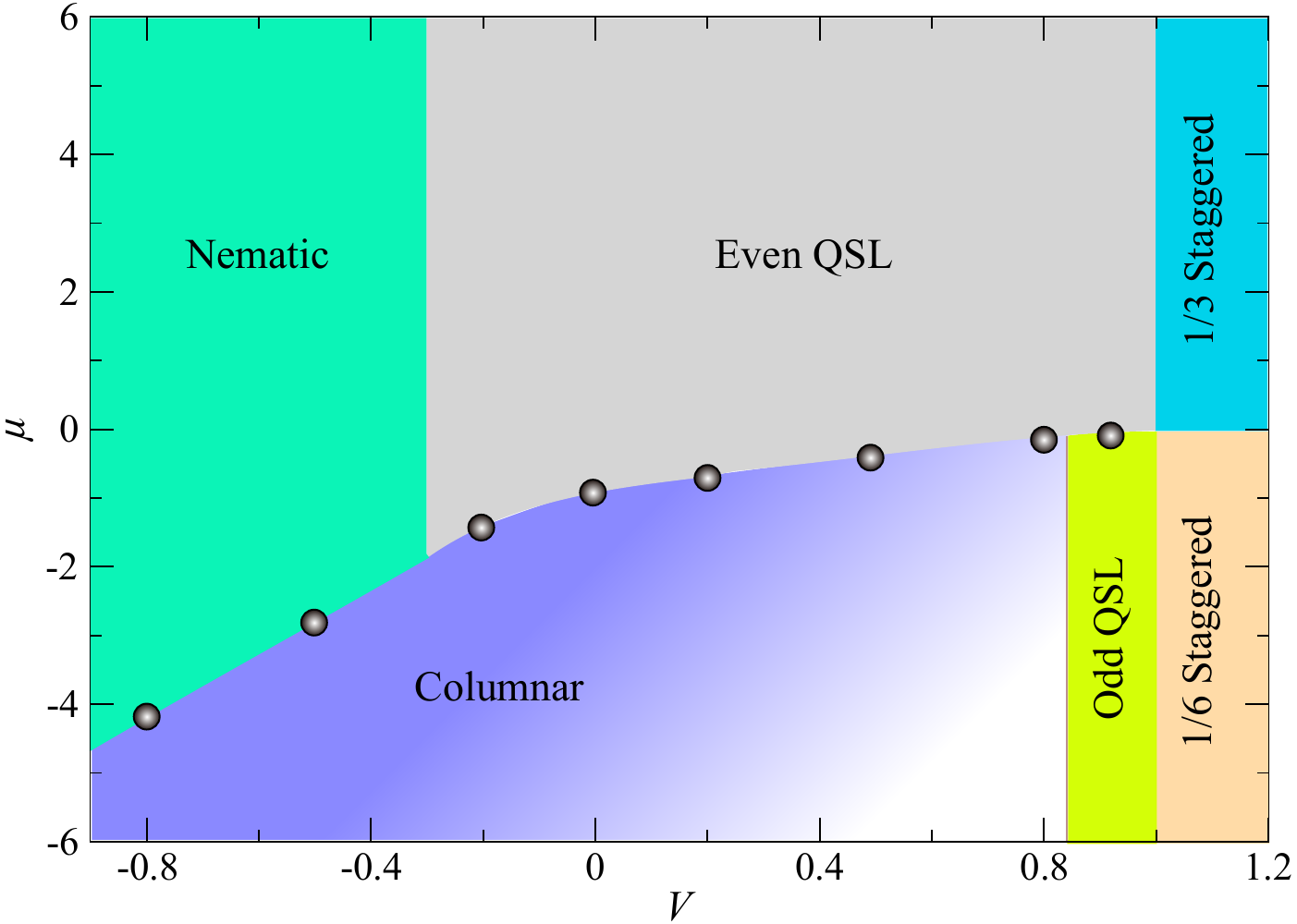}
\caption{\textbf{Phase diagram of $h=0$ case.} The phase diagram, spanned by the $V$ and $\mu$ axes, obtained from QMC simulations at $h=0$. The QSL--nematic and QSL--columnar transitions are continuous, and the QSL--staggered transition is first-order. In the limit of exactly one dimer per site, a $\sqrt{12}\times\sqrt{12}$ valence bond solid (VBS) phase is known to exist between the odd QSL and the columnar phase. However, it is nearly degenerate with the columnar phase over a large region in our simulations---especially for larger sizes---and we depict this schematically by using a lighter shading for the columnar phase near the odd QSL.} 
\label{SF2}
\end{figure*}
\vspace{\baselineskip}
\noindent {\bf Supplementary Note 2: Phase diagram at $h=0$.}

Besides the phase diagram with $h=0.4$ presented in the main text, we also study the phase diagram at $h=0$ described by the Hamiltonian,
\begin{eqnarray}
  H=&-t&\sum_r \left(
  \left|\rhombV\right>\left<\rhombH\right| + \mbox{h.c.}
  \right)  +V\sum_r\left(
  \left|\rhombV\right>\left<\rhombV\right|+\left|\rhombH\right>\left<\rhombH\right|
  \right) -\mu \sum_l \left(\left|\onedimer\right>\left<\onedimer\right|\right),
\label{SM:eq1}
\end{eqnarray}
where the sum on $r$ runs over all plaquettes (rhombi) including the three possible
orientations. The kinetic term $t$, the potential term $V$, and the chemical potential $\mu$ are the same as in Eq.(1) of the main text; we set $t=1$ as the unit of energy. As in the finite-$h$ case, we impose a soft constraint requiring either one or two nearest-neighbor dimer(s) per site.

Since the Hamiltonian cannot flip single dimers on a bond in the absence of an $h$ term, the dimer filling becomes a conserved quantity. First, we note that with $V$\,$=$\,$1$ but varying $\mu$, the system is described by the Rokhsar-Kivelson (RK) wavefunction of an equal superposition of dimer coverings~\cite{RK88}, thus forming a QSL ground state. Moreover, the ground-state energy of the Hamiltonian [Supplementary Eq.~\eqref{SM:eq1}] at the RK point, without the $\mu$ term, is identically zero for a fixed filling $\rho$. A nonzero $\mu$ trivially makes the ground state favor one (two) dimers per site when $\mu<0$ ($>0$). Similarly, the $1/6$- and $1/3$-filling staggered phases are degenerate when $\mu=0$. Hence, the (first-order) phase transition line separating these two staggered phases remains at $\mu=0$.

For the phase boundaries determined by fixing $\mu$ and varying $V$, if the phases have the same filling, the chemical potential $\mu$ will not change their energy difference but just impart equal energy shifts. Therefore, in the vertical direction, the phase boundaries are just straight lines. Likewise, the horizontal phase boundaries, which separate phases of two different fillings, are also straight. Together, they give rise to the entire phase diagram in Supplementary Fig.~\ref{SF2}.

In the limit of large, negative $V$, the kinetic term becomes irrelevant. The Hamiltonian then becomes a classical one with competing terms set by $\mu$ and $V$. It is not hard to see that the phase transition line between the columnar and nematic phases is simply given by $V=\mu/3$ in this classical limit.

\begin{figure}[thp!]
\centering
    \includegraphics[width=\columnwidth]{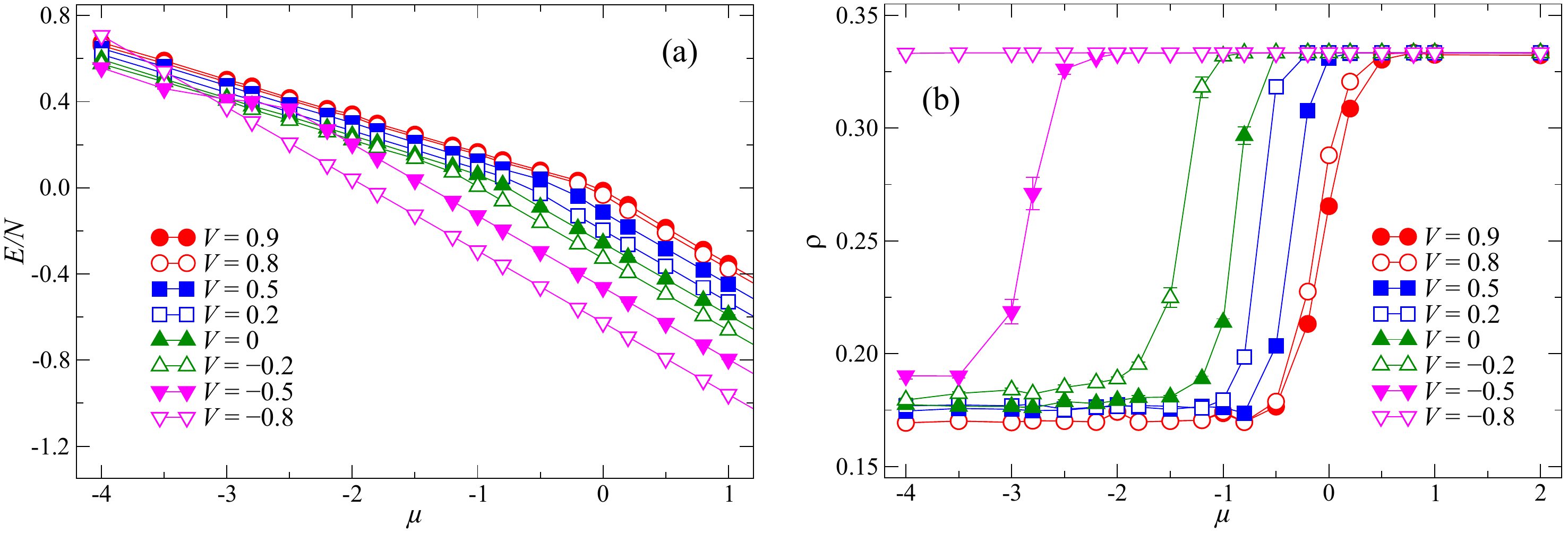}
    \caption{\textbf{The relation of Energy/filling and $\mu$ of different $V$.} (a) Energy density $E/N$, and (b) dimer filling $\rho$ for different $V$ while scanning $\mu$ at a fixed size $L=12$. The phase transition line progressively shifts towards more negative values of $\mu$ as $V$ is decreased.}
    \label{SF3}
\end{figure}

\begin{figure}[thp!]
\centering
    \includegraphics[width=\columnwidth]{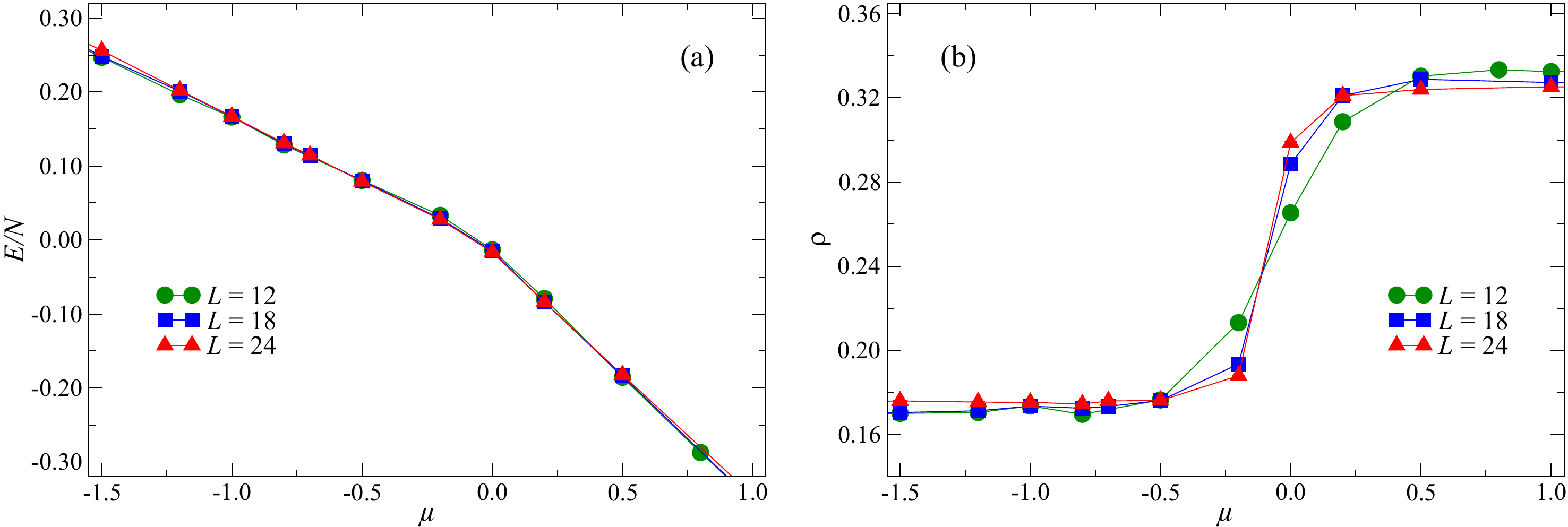}
    \caption{\textbf{The relation of Energy/filling and $\mu$ of different $L$ at $V=0.9$.} (a) The energy densities for different system sizes exhibit similar behaviours as a function of $\mu$. (b) The dimer filling $\rho$ becomes sharper while the size increases. This data confirms an obvious first-order phase transition between the odd and even QSLs.}
    \label{SF4}
\end{figure}

\begin{figure}[thp]
\centering
\includegraphics[width=\columnwidth]{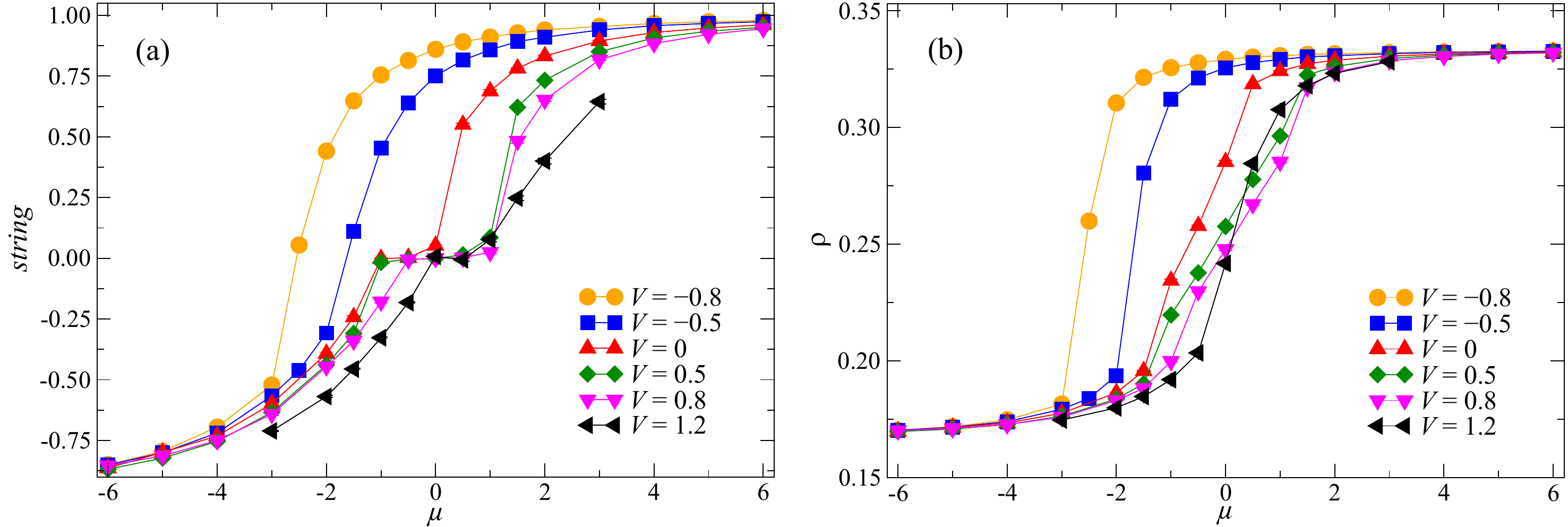}
\caption{\textbf{The relation of string/filling and $\mu$ of different $V$.} (a) The string operator $\langle string \rangle$ and (b) the dimer filling $\rho$ as a function of $\mu$ for different values of $h=0.4$, and $V=0.8$, $0.5$, $0$, $-0.5$, $-0.8$ in a system of size $L=16$. The system size simulated for $V=1.2$ is chosen to be $L=12$ due to the unit cell of the staggered order.}
    \label{SF5}
\end{figure}

To understand the phase diagram at a quantitative level, we simulate the model at a fixed system size of $L=12$. Clear first-order phase transitions arise between the columnar and nematic, the columnar and even QSL, and the odd and even QSL phases, as can be seen from the energy density $E$ and the dimer filling $\rho$ (Supplementary Fig.~\ref{SF3}). We have also studied the system-size dependence of the odd and even QSL transition at $V=0.9$ with $L=12,18$ and $24$. As shown in Supplementary Fig.~\ref{SF4}, the first-order phase transition also becomes more obvious as the system size increases, and there is no PM phase between the two QSLs unlike for the finite-$h$ cases, which we now turn to discuss.

\vspace{\baselineskip}
\noindent {\bf Supplementary Note 3: Additional data for the phase diagram at $h=0.4$.}

In this section, we provide the detailed data used to construct the phase diagram sketched in Fig.~1 of the main text.

First, the string operator $\langle string \rangle$ and the dimer filling $\rho$ are used to distinguish between the PM and other phases. In Supplementary Fig.~\ref{SF5}, we present these observables as a function of $\mu$ for fixed $V=0.8, 0.5, 0, -0.5, -0.8$. They clearly show the vanishing of the PM region as $V$ is varied from positive to negative values. At $V=-0.8$, a clear first-order transition between the nematic phase (with $\rho\sim 1/3$) and the columnar phase (with $\rho \sim 1/6$) is manifest.

Moreover, starting from $h=0$, we found that the extent of the PM phase increases with increasing $h$. To demonstrate this behavior, we measure the string operator at $h=$0.1, 0.2, 0.4 for a fixed $V=$0.9 and varying $\mu$. The results are shown in Supplementary Fig.~\ref{SF6}. It is clear that at $h=0.1$, $\langle string \rangle$ jumps from $-1$ to $1$ at $\mu \sim 0$, close to the odd QSL to even QSL first-order transition observed at $h=0$ for the phase diagram of Supplementary Fig.~\ref{SF2} in the previous section. However, as $h$ increases, to $0.2$ and $0.4$, the intermediate PM phase, with $\langle string \rangle = 0$, separating the two QSLs becomes clearer.

\begin{figure}[t]
\centering
    \includegraphics[width=0.5\columnwidth]{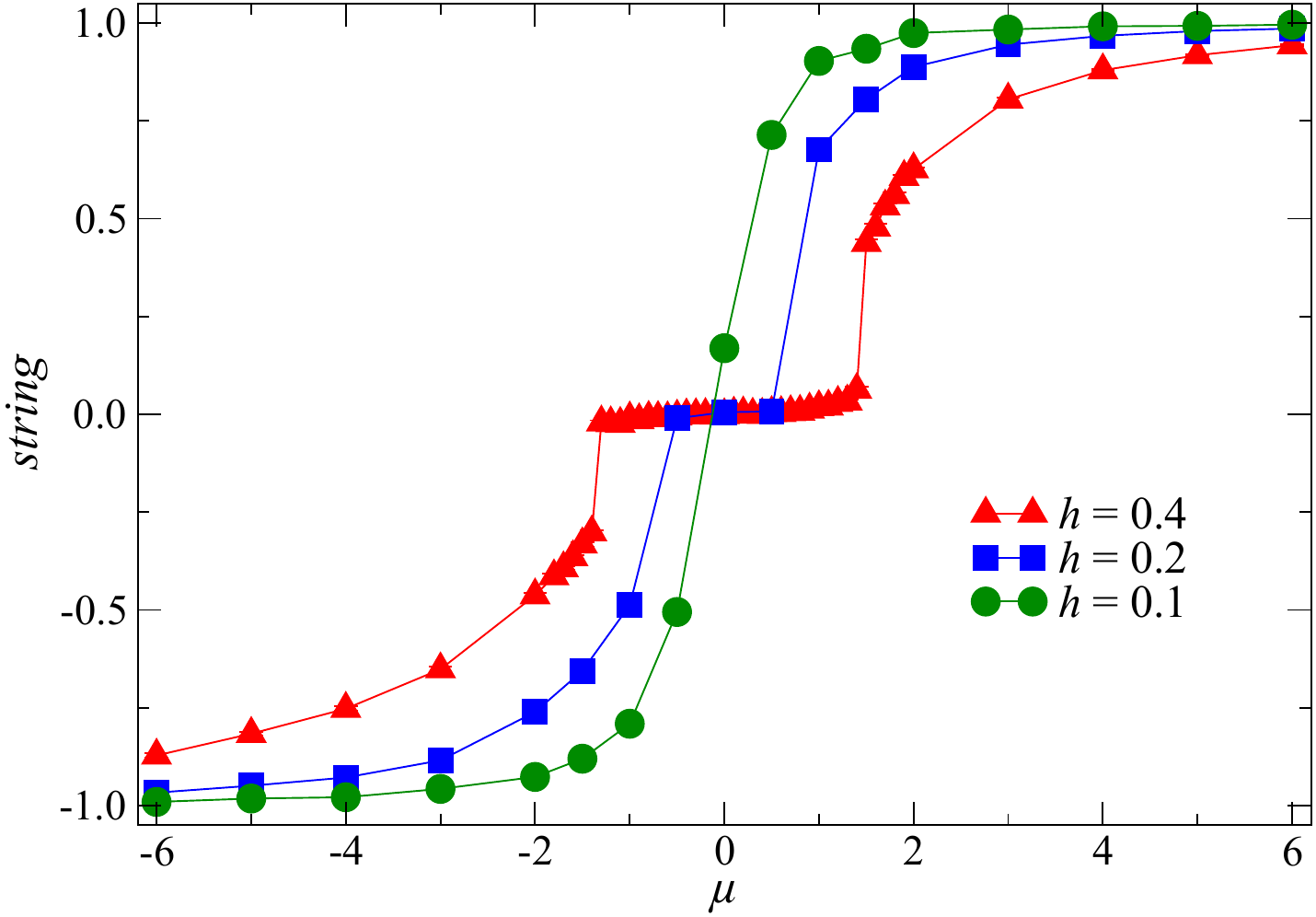}
    \caption{\textbf{The relation of string and $\mu$ of different $h$.} The string operator as function of $\mu$ for $h=0.1$, $0.2$, $0.4$, with $V=0.9$ and a system size of $L=16$. It shows the region of PM becomes larger when the $h$ increasing.}
    \label{SF6}
\end{figure}

\begin{figure}[b]
\centering
\includegraphics[width=\columnwidth]{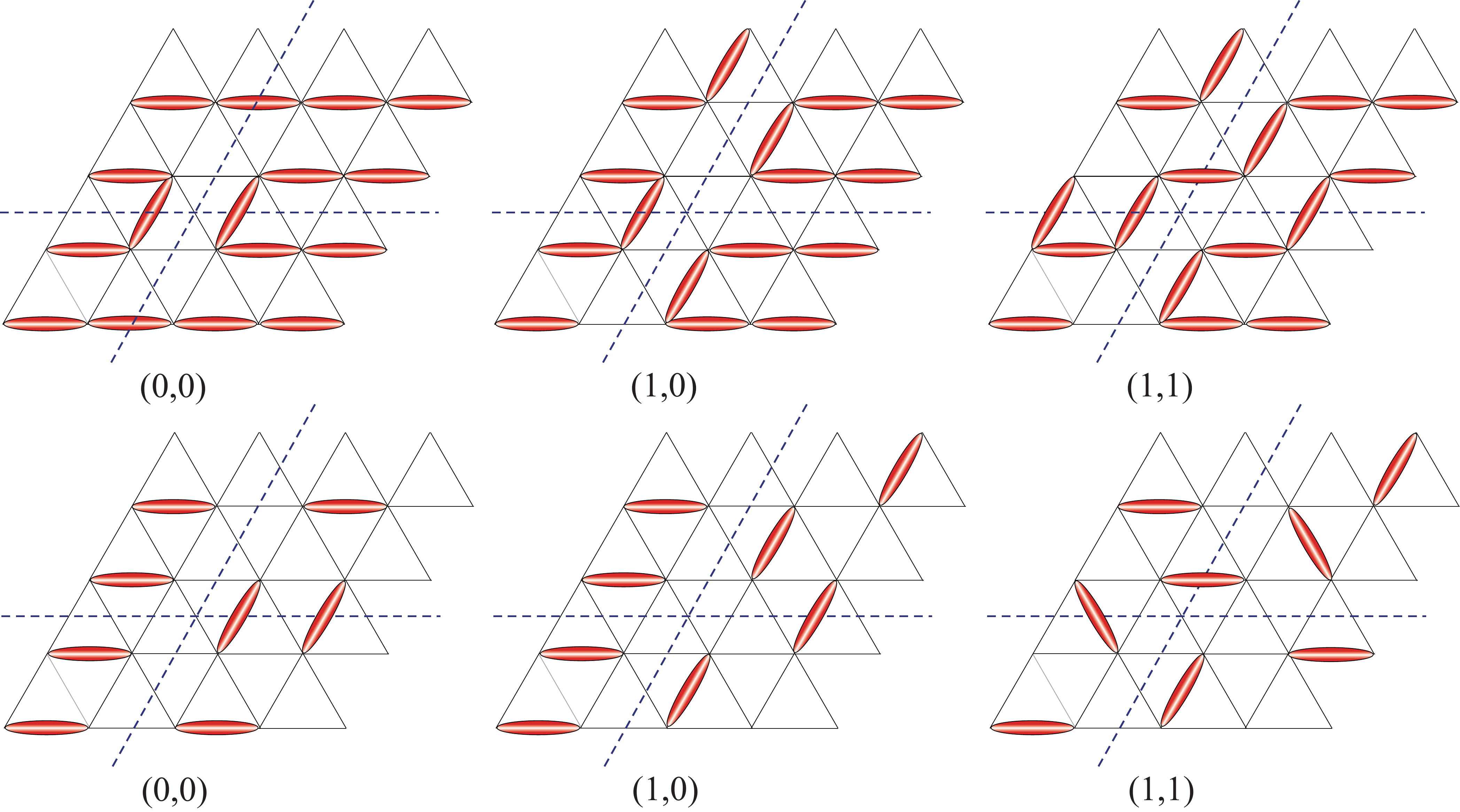}
\caption{\textbf{The definition of winding number.} The winding number is well-defined on a lattice with periodic boundary conditions. The number of dimers cut (modulo $2$) along the two directions, as shown by the dashed lines, yields the winding number $(x,y)$. The upper (lower) row arrays a few examples of dimer configurations with different winding numbers in the $1$- ($2$)-dimer-per-site limit.}
\label{SF7}
\end{figure}

Going back to the $h=0.4$ phase diagram, on the positive (negative) $\mu$ side, one finds the $1/3$ ($1/6$) staggered phase and the even (odd) QSL near $V=1$. As $V$ approaches negative values, the phase transition between the QSL and the valence bond solid (VBS) phase is proposed to be continuous and in the O(3)$^*$ (O(4)$^*$) universality class~\cite{roychowdhury2015z,ZY2020} with large anomalous dimension exponents~\cite{Isakov2012,YCWang2018}. However, we note that the precise nature of this topological transition is still largely unknown; here, we use the energy differences between different sectors to roughly estimate the position of the phase transition, following previous examples~\cite{Ralko2005}. The basic idea is that the VBS state belongs to the (0,0) sector of dimer coverings on the torus geometry, and the other sectors---such as (1,1) with two topological defects along the $x,y$ axes---will cost more energy~\cite{ZY2020improved,zhou2020quantumstring,zhou2020quantum,yan2022targeting} as they create  global domain walls. On the other hand, these sectors are topologically degenerate in the $\mathbb{Z}_2$ QSL phase. In the QMC simulation, we can prepare the dimer states in these different sectors, as shown in Supplementary Fig.~\ref{SF7}, and monitor their energy difference as we scan through the transition from the VBS to the $\mathbb{Z}_2$ QSL phase. As the system size increases, the energy difference vanishes inside the QSL and saturates to a finite value (note that the difference is scaled such that it is intensive) inside the VBS phase. The boundary between these two behaviors is then taken to be the boundary of the two phases.

\begin{figure}[b]
\centering
\includegraphics[width=\columnwidth]{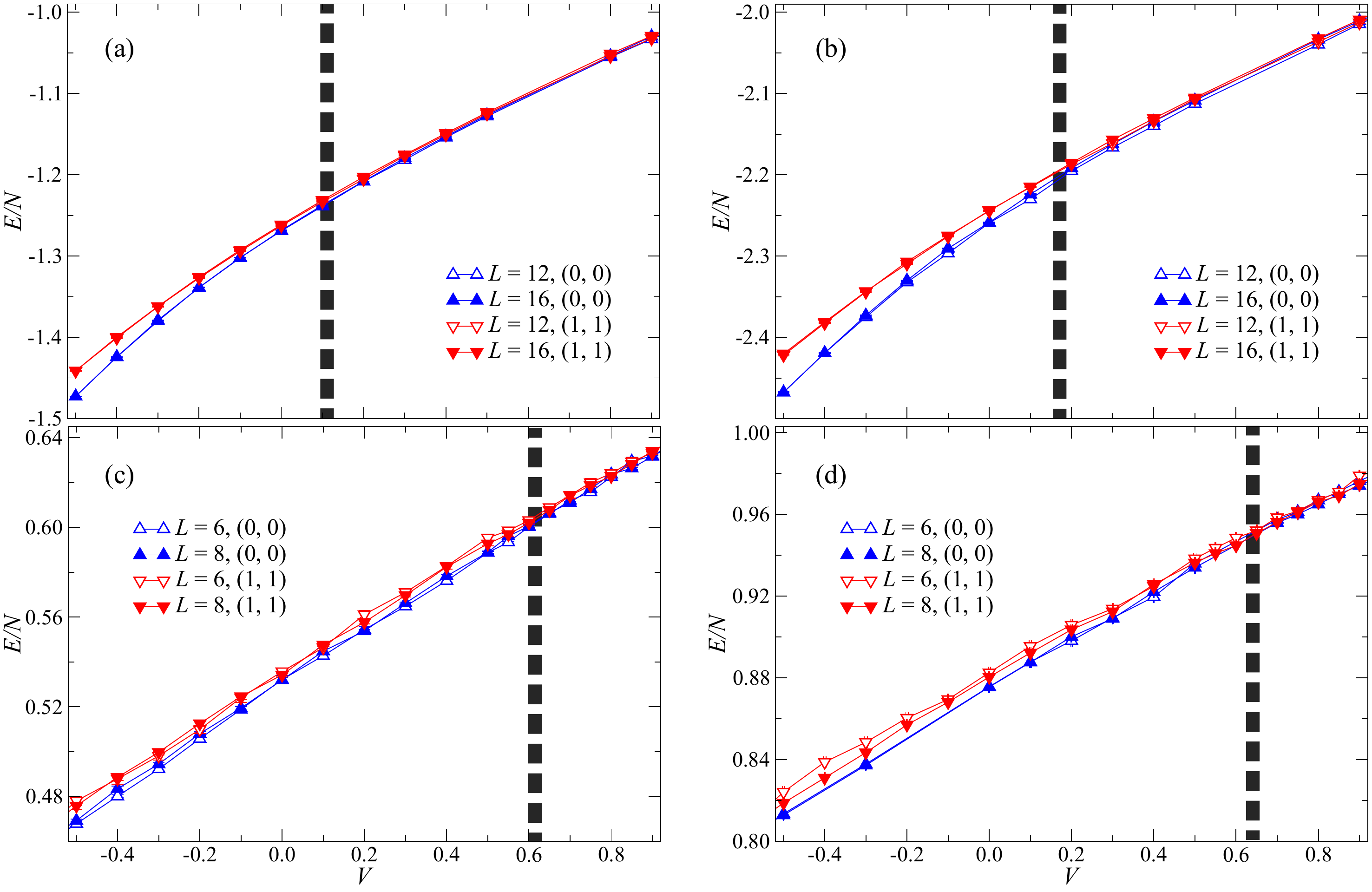}
\caption{\textbf{The energy difference of topological sectors.} The energy difference per site in different topological sectors as we scan $V$ through the nematic VBS to even $\mathbb{Z}_2$ QSL phase transition for $\mu=3$ (a), and $\mu=6$ (b). The two sectors $(0,0)$ and $(1,1)$ have a finite energy difference in the nematic VBS and become degenerate in the QSL, especially as the system size increases. The vertical dashed lines denote the transition point within our simulation resolution. A similar approach is used to distinguish the columnar VBS and the odd QSL roughly at $\mu=-4$ (c), and $\mu=-6$ (d).}
\label{SF8}
\end{figure}

The results of such an analysis are shown in Supplementary Fig.~\ref{SF8}, where we scan $V$ for two different chemical potentials, $\mu=3$ and $\mu=6$. As the linear system size increases from $L=12$ to $16$, the energy difference ($E/N$ on the $y$ axis) between the topological sectors $(0,0)$ and $(1,1)$ remains nonzero in the nematic phase but becomes vanishingly small in the $\mathbb{Z}_2$ QSL phase. We use the vertical dashed lines to denote the transition points determined in this fashion, and these are the phase boundaries presented in the phase diagram of the main text.

\begin{figure}[thp]
\centering
\includegraphics[width=\columnwidth]{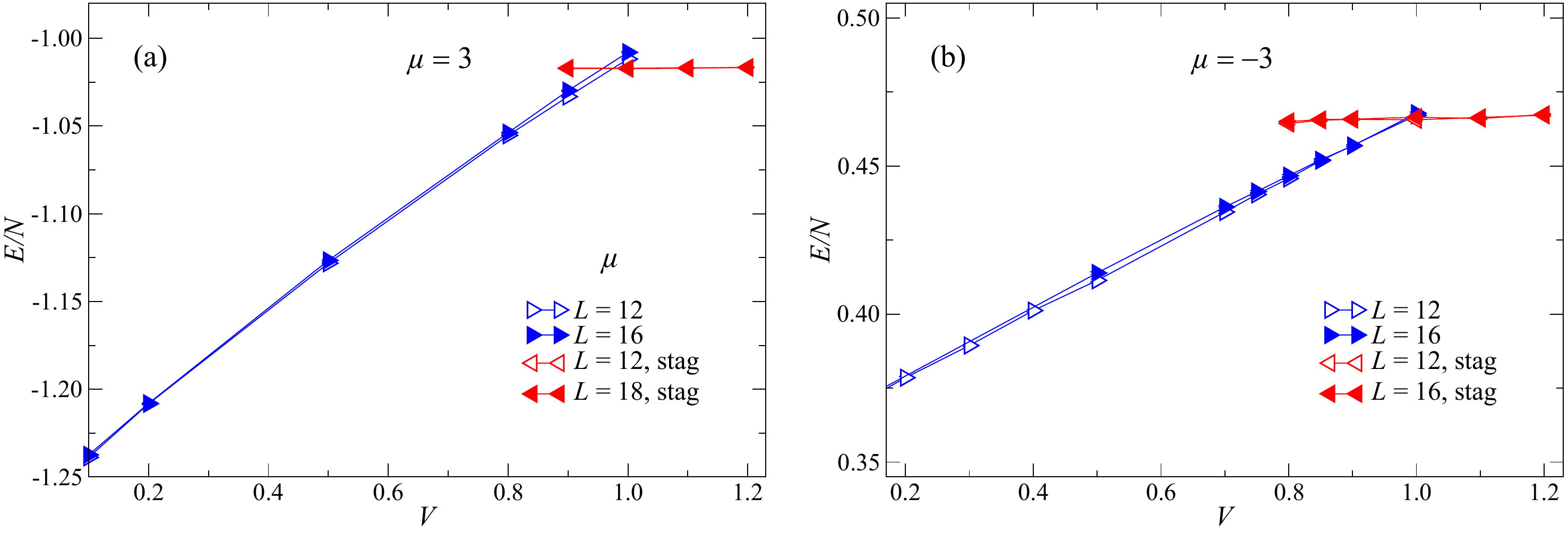}
\caption{\textbf{Energy cross for the first order phase transition.} The staggered phases belongs to different topological sectors than the QSL phases. The staggered-to-QSL phase transitions are found to be first-order by examining the energy density curves at (a) $\mu=3$ and (b) $\mu=-3$. Clearly, the lines for the staggered and QSL phases cross at $\sim0.98$, which is close to the $V=1$ QSL--staggered phase transition point of the original dimer models (cf. Supplementary Fig.~\ref{SF2}).}
    \label{SF9}
\end{figure}

Lastly, we discuss the phase transition between the $1/3$- or $1/6$-staggered VBS phases at large positive $V$ and the corresponding QSL phases. In the phase diagram without $h$ (Supplementary Fig.~\ref{SF2}), the staggered phases at both $1/3$ and $1/6$ fillings are separated from their proximate QSLs by first-order phase transitions. This behavior persists in the case of $h=0.4$ as well, as illustrated by Supplementary Fig.~\ref{SF9}. Here, we plot the energy densities with the initial configuration chosen to be either the QSL or the staggered state, and find that the two different energy curves cross each other at the first-order transition point $\sim0.98$ for two values of $\mu$. The phase boundaries in the main text are determined by this procedure.

%